\documentclass[%
 reprint,
 amsmath,amssymb,
 aps,
]{revtex4-2}

\usepackage{graphicx}
\usepackage{xcolor}
\usepackage{soul}
\usepackage{dcolumn}
\usepackage{bm}
\usepackage[colorlinks=true, allcolors=blue]{hyperref}

\usepackage{amsmath} 
\usepackage[utf8]{inputenc} 
\usepackage{physics}
\usepackage[letterpaper, margin=1in]{geometry}

\newcommand{\R}{\mathbb R}
\newcommand{\Z}{\mathbb Z}

\newcommand{\C}{\mathbb C}

\begin{document}

\preprint{APS/123-QED}

\title{Bifurcation delay and front propagation in the real Ginzburg-Landau equation on a time-dependent domain}

\author{Troy Tsubota$^1$}
\email{tktsubota@berkeley.edu}
\author{Chang Liu$^{1,2}$}
\email{chang\_liu@uconn.edu}
\author{Benjamin Foster$^1$}
\email{ben\_foster@berkeley.edu}
\author{Edgar Knobloch$^1$}
\email{knobloch@berkeley.edu}
\affiliation{$^1$Department of Physics, University of California at Berkeley, Berkeley, California 94720, USA}
\affiliation{$^2$School of Mechanical, Aerospace, and Manufacturing Engineering, University of Connecticut, Storrs, Connecticut 06269, USA}
\date{April 12, 2024}

\begin{abstract}
This work analyzes bifurcation delay and front propagation in the one-dimensional real Ginzburg-Landau equation with periodic boundary conditions on isotropically growing or shrinking domains. First, we obtain closed-form expressions for the delay of primary bifurcations on a growing domain and show that the additional domain growth before the appearance of a pattern is independent of the growth time scale. We also quantify primary bifurcation delay on a shrinking domain; in contrast with a growing domain, the time scale of domain compression is reflected in the additional compression before the pattern decays. For secondary bifurcations such as the Eckhaus instability, we obtain a lower bound on the delay of phase slips due to a time-dependent domain. We also construct a heuristic model to classify regimes with arrested phase slips, i.e. phase slips that fail to develop. Then, we study how propagating fronts are influenced by a time-dependent domain. We identify three types of pulled fronts: homogeneous, pattern-spreading, and Eckhaus fronts. By following the linear dynamics, we derive expressions for the velocity and profile of homogeneous fronts on a time-dependent domain. We also derive the natural ``asymptotic'' velocity and front profile and show that these deviate from predictions based on the marginal stability criterion familiar from fixed domain theory. This difference arises because the time-dependence of the domain lifts the degeneracy of the spatial eigenvalues associated with speed selection and represents a fundamental distinction from the fixed domain theory that we verify using direct numerical simulations. The effect of a growing domain on pattern-spreading and Eckhaus front velocities is inspected qualitatively and found to be similar to that of homogeneous fronts. These more complex fronts can also experience delayed onset. Lastly, we show that dilution---an effect present when the order parameter is conserved---increases bifurcation delay and amplifies changes in the homogeneous front velocity on time-dependent domains. The study provides  general insight into the effects of domain growth on pattern onset, pattern transitions, and front propagation in systems across different scientific fields. 
\end{abstract}

\maketitle

\section{Introduction}\label{sec:introduction}

Pattern formation on a time-dependent domain arises in many systems across biology and physics \cite{turing_chemical_1952, crampin_reaction_1999, madzvamuse_stability_2010, knobloch_problems_2015}.
Key examples include the expanding crown in the drop-splash problem \cite{krechetnikov_linear_2011}, structure formation in an expanding universe  \cite{peebles_large-scale_2020}, and a variety of reaction-diffusion problems on growing domains \cite{gierer_theory_1972,meinhardt_pattern_1992, crampin_reaction_1999,crampin_pattern_2002,plaza_effect_2004,madzvamuse_stability_2010, goriely_mathematics_2017,kim_pattern_2020,ben_tahar_turing_2023}. Other problems, ranging from quantum mechanics \cite{duca_schrodinger_2021, duca_control_2023} to control theory \cite{dai_closed_2023}, have also been studied on time-dependent domains. For a review of such problems, see \cite{knobloch_problems_2015}.

When the domain size becomes time-dependent, new phenomena emerge in the dynamic pattern formation process. For example, the stability of pattern states can change but on a time-dependent domain these changes occur at stronger forcing than that predicted by quasi-static changes in the domain size---a phenomenon referred to as bifurcation delay \cite{benoit_dynamic_1991}. A better understanding of bifurcation delay can provide additional physical insights into climate tipping points \cite{ashwin_tipping_2012}, oscillator networks \cite{premraj_experimental_2016, premraj_effect_2019, varshney_bifurcation_2020, talla_mbe_study_2020}, neuron firing \cite{rinzel_threshold_1988}, and many other applications \cite{ahlers_amplitude_1981, laplante_jump_1991,strizhak_slow_1996,de_maesschalck_canards_2009,vasil_dynamic_2011, premraj_control_2017,yu_delayed_2023}. Bifurcation delay in systems described by nonautonomous ordinary differential equations has been extensively studied \cite{mandel_slow_1987, baer_slow_1989, lythe_domain_1996, haberman_slow_2001, diminnie_slow_2002, ng_slow_2003, neishtadt_stability_2009, park_slow_2011, han_delayed_2014}, but similar phenomena in spatially extended systems described by partial 
differential equations are much less understood \cite{knobloch_stability_2014, knobloch_problems_2015, krechetnikov_stability_2017} and no explicit formulas are available outside of the adiabatic regime.


The phenomenon of front propagation, i.e., the motion of an interface between stable and unstable states or between two different stable states, is also of interest. The simplest example of a front occurs when a stable, spatially homogeneous state propagates into an unstable homogeneous state. Fronts of this type generally travel with a velocity that can be computed from linearized dynamics alone, although  nonlinearities can amplify the propagation speed \cite{van_saarloos_front_2003}. Fronts arise in many systems in fluid, chemical, and biological environments such as vortex fronts in Taylor-Couette flow \cite{ahlers_vortex-front_1983}, pearling instabilities of lipid bilayers \cite{powers_propagation_1998}, and healing of epidermal wounds \cite{maini_travelling_2004}. For a review of the theory and applications of fronts on a fixed domain, see \cite{van_saarloos_front_2003}. Front propagation gains additional complexity on time-varying domains or in systems with evolving parameters \cite{stoop_defect_2018, rietkerk_evasion_2021, goh_fronts_2022, goh_growing_2023}. It is currently unclear how a time-dependent domain affects front propagation.


In conserved systems, bifurcation delay and front propagation are affected by the presence of a conserved quantity. As the domain expands, the concentration of this quantity is reduced, yielding a local growth-dependent effect known as dilution. A shrinking domain has the opposite effect of increasing the concentration.
For one-dimensional isotropic domain growth, the effect of dilution can be transformed into a stabilizing time-dependent bifurcation parameter \cite{krechetnikov_stability_2017}.

To understand the properties of these phenomena on a time-dependent domain, we study the one-dimensional (1D) real Ginzburg-Landau equation (RGLE) model. The RGLE captures many of the phenomenological properties of pattern formation with minimal complexity. Furthermore, the RGLE provides a quantitative description of amplitude modulation of more complex patterns near the onset of instability, such as those found in Rayleigh-B{\'e}nard convection and Taylor-Couette flow \cite{segel_distant_1969, newell_finite_1969, ahlers_amplitude_1981, ahlers_wavenumber_1986, cross_pattern_1993}. The dynamics of the RGLE on a fixed domain are well-understood \cite{kramer_effects_1984, tuckerman_bifurcation_1990, eckmann_front_1993, cross_pattern_1993, bridges_instability_1994, doelman_instability_1995, eckmann_phase_1995, cross_pattern_2009, knobloch_problems_2015}. However, existing theory is often not generalizable to a time-dependent domain which incorporates nonautonomous effects on an arbitrary time scale.

The RGLE on a time-dependent domain was previously studied in \cite{knobloch_stability_2014, knobloch_problems_2015}, focusing on slowly changing domains relative to the intrinsic time scale. In this work, an expression for the bifurcation delay of a phase slip under slow domain changes was derived based on a time-dependent diffusion equation for the spatial phase \cite{knobloch_stability_2014}. A local theory for phase slips was also developed based on the regularity properties of parabolic partial differential equations (PDEs) to show how a growing (shrinking) domain increases (decreases) the time to phase slip \cite{knobloch_problems_2015}. This work also showed that the nonlinear evolution of the Eckhaus instability can be described with a nonlinear porous-medium type equation \cite{knobloch_stability_2014}.

The present work analyzes bifurcation delay and front propagation in the RGLE on isotropically growing or shrinking domains but relaxes the assumption of slow time variation through both theoretical analysis and extensive direct numerical simulations. Additional phenomena arise in models with an intrinsic wave number such as those motivated by apical growth \cite{goh_universal_2016}. First, we obtain explicit formulas for the bifurcation delay of primary bifurcations in both growing and shrinking domains of varying time scales. For secondary bifurcation delay, we construct an energy bound on perturbations on a shrinking domain and classify regimes containing arrested phase slips, or phase slips that fail to develop due to re-stabilization on a growing domain. Next, we identify several types of fronts in the RGLE and explain how their dynamics change on a time-dependent domain. For homogeneous fronts, we derive a linear spreading velocity and a natural ``asymptotic'' velocity which compares well with direct numerical simulations. We also analyze the time evolution of the nonlinear front profile and its connection to the front velocity. Lastly, we show how dilution increases bifurcation delay and amplifies changes in homogeneous front velocity.

In Section \ref{sec:formulation}, we formulate the RGLE on a time-dependent domain. In Sections \ref{sec:bifurcation_delay} and \ref{sec:fronts}, we analyze the effect of a time-dependent domain on bifurcation delay and front propagation, respectively. In Section \ref{sec:dilution}, we explain the role of dilution. In Section \ref{sec:discussion}, we summarize our results and suggest directions for future study.

\section{Real Ginzburg-Landau Equation on a Time-Dependent Domain}\label{sec:formulation}

\begin{figure*}
\centering
\includegraphics[width=\textwidth]{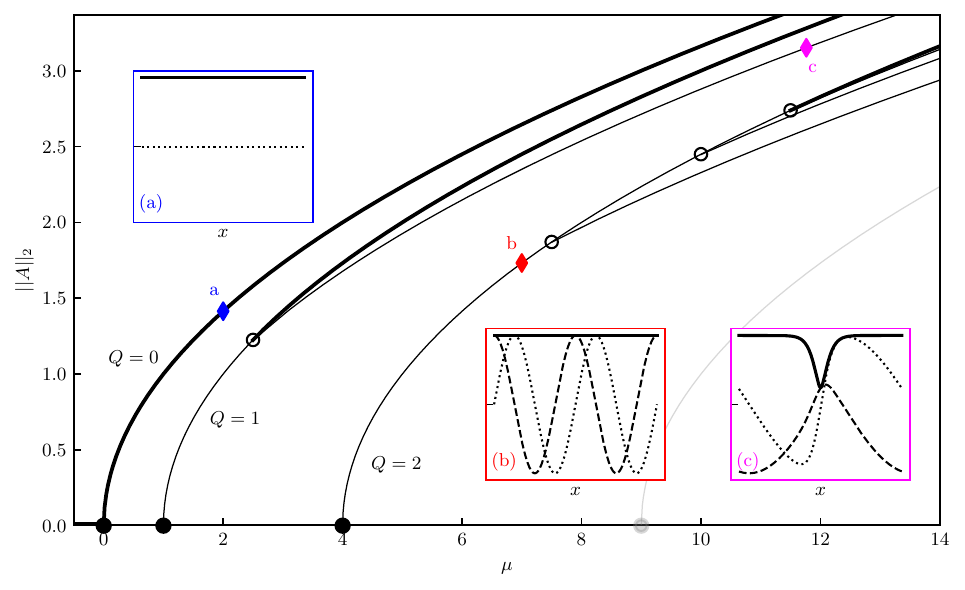} 
\caption{Bifurcation diagram showing $||A||_2$ vs. $\mu$ for Eq. \eqref{eq:RGLE_no_dilution} with periodic boundary conditions, $\Lambda=2\pi$ and $L=1$, where $||A||_2 = \sqrt{\frac{1}{\Lambda}\int_{0}^{\Lambda}|A|^2 \, dx}$ is the (normalized) $L^2$ norm. Solid lines represent unstable (thin) and stable (thick) stationary states. Filled circles denote primary bifurcations at $\mu=Q^2$ for each wave number $Q$, and open circles denote secondary bifurcations at $\mu=3Q^2-\frac{1}{2}k^2$ for each integer $k\in(0,2Q)$. In the insets (a), (b) and (c), we show sample profiles for the three key stationary states: solid, dashed, and dotted lines denote $|A|, \ \Re A, $ and $\Im A$, respectively, and the $y$-ticks denote zero. The homogeneous state (a) is given by $A=\sqrt{\mu}$. The phase-winding pattern states (b) have the form $A=\sqrt{\mu-Q^2}e^{iQx}$. The mixed modes (c) determine the basin of attraction of the pattern states. The corresponding locations of each inset profile in the bifurcation diagram are marked with diamonds.}
\label{fig:mu_bifurcation_diagram}
\centering
\end{figure*}

We consider a complex amplitude $A$ described by the 1D RGLE on an isotropically growing domain $x\in [0, \Lambda L(t)]$ with periodic boundary conditions:
\begin{equation}\label{eq:RGLE_no_conservation}
    A_t + \underbrace{\frac{\dot{L}(t)}{L(t)}xA_x}_{\text{advection}} = \mu A + A_{xx} - |A|^2A.
\end{equation}
Here, subscripts denote partial derivatives and overdots denote total time derivatives, while $L(t)$ is the dimensionless growth parameter; we prescribe $L(0)=1$ so that $\Lambda$ is the initial domain size. The RGLE is the lowest-order model that retains phase invariance $A\mapsto Ae^{i\phi}$ for arbitrary phase $\phi$ and separate parity symmetries $A\mapsto \overline{A}$ and $x \mapsto -x$ present in many physical systems \cite{cross_pattern_2009}. If we treat the RGLE as a model pattern-forming equation in its own right, we may impose the additional requirement that $\int_0^{\Lambda L(t)}A(x,t)\,dx$ must be conserved by the left hand side, leading to
\begin{equation}\label{eq:RGLE_eulerian}
    A_t + \frac{\dot{L}(t)}{L(t)}xA_x + \underbrace{\frac{\dot{L}(t)}{L(t)}A}_{\text{dilution}} = \mu A + A_{xx} - |A|^2A.
\end{equation}

Existing analytical techniques and numerical methods for PDEs require a fixed domain. Therefore, we use the change of variable $\xi = x/L$ to obtain
\begin{equation}\label{eq:RGLE}
    A_t + \frac{\dot{L}(t)}{L(t)} A= \mu A + \frac{1}{L(t)^2}A_{\xi\xi} - |A|^2A,
\end{equation}
where $\xi\in[0,\Lambda]$ is the (Lagrangian) fixed-domain spatial coordinate. Note that the advection term is eliminated in this frame; see Appendix \ref{app:dilution}. To isolate the effect of dilution, we initially neglect the dilution term in \eqref{eq:RGLE} and study the equation
\begin{equation}\label{eq:RGLE_no_dilution}
    A_t = \mu A + \frac{1}{L(t)^2}A_{\xi\xi} - |A|^2A.
\end{equation}
This is the relevant regime when treating the RGLE as an amplitude equation for pattern-forming systems. The dilution term is reintroduced in Section \ref{sec:dilution} once this simpler system is analyzed. 

The RGLE possesses several classes of stationary states summarized in the bifurcation diagram in Fig.~\ref{fig:mu_bifurcation_diagram} for $L=1$ and $\Lambda = 2\pi$. These states are defined up to a constant phase shift arising from translation invariance and the use of periodic boundary conditions. The supercritical bifurcations along the trivial branch $A=0$ are referred to as primary bifurcations and these generate pattern states. The restabilizing subcritical bifurcations along the pattern branches are referred to as secondary bifurcations, and these are responsible for the presence of unstable mixed mode solutions.

\begin{figure*}
\centering
\includegraphics[width=\textwidth]{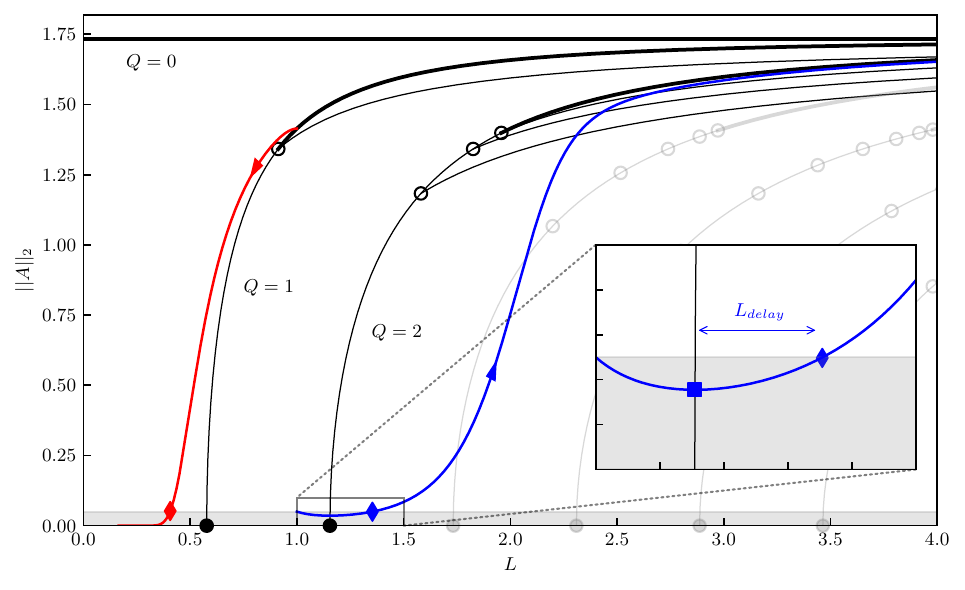} 
\caption{Bifurcation diagram showing  $||A||_2$ vs. $L$ for  $\mu = 3$. Solution stability and bifurcation types are denoted as in Fig.~\ref{fig:mu_bifurcation_diagram}. The primary bifurcations occur at $L=Q/\sqrt{\mu}$, $Q\in \mathbb{Z}^+$, and secondary bifurcations occur at $L=\sqrt{(3Q^2-\frac{1}{2}k^2)/\mu}$ for each $k\in(0,2Q)$ with $k\in \mathbb{Z}^+$. The trajectory for a $Q=2$ state on a growing domain $L(t)=e^{0.2t}$ is shown in blue with an upward arrow, and the trajectory for a $Q=1$ state on a shrinking domain $L(t)=e^{-0.2t}$ is displayed in red with a downward arrow. The shaded region denotes the neighborhood around the trivial state for which delay is calculated; diamonds represent $L_{exit}$ (blue, right) and $L_{enter}$ (red, left) for the growing and shrinking domain, respectively. In the inset, the square denotes the turnaround point $L_*$. The additional domain growth $L_{delay}$ denotes bifurcation delay.}
\label{fig:L_bifurcation_diagram}
\centering
\end{figure*}

To relate the properties of Eq.~\eqref{eq:RGLE_no_dilution} to the time-independent problem, it is often useful to freeze the time dependence of $L(t)$ and treat it as a bifurcation parameter of the system:
\begin{equation}\label{eq:RGLE_no_dilution_frozen}
    A_t = \mu A + \frac{1}{L^2}A_{\xi\xi} - |A|^2A.
\end{equation}
This allows us to obtain the bifurcation diagram of the RGLE with respect to $L$ as shown in Fig.~\ref{fig:L_bifurcation_diagram} for fixed $\mu=3$. As a system parameter, the domain size $L$ behaves similarly to $\mu$: pattern states become more stable as $L$ increases. However, this is a quasi-static view and does not capture all time-dependent effects of $L(t)$. 

\section{Bifurcation Delay}\label{sec:bifurcation_delay}

\subsection{Primary Bifurcation Delay}

Under quasi-static variation of $L$, supercritical primary bifurcations from the trivial state to pattern states of wave number $Q$ and amplitude $\sqrt{\mu-Q^2/L^2}$ occur at $L=Q/\sqrt{\mu}$ as shown in Fig.~\ref{fig:L_bifurcation_diagram}. On a growing domain, the onset of these pattern states can be delayed beyond the primary bifurcation. Conversely, on a shrinking domain, pattern states can exhibit delayed decay. We aim to explicitly measure these delay effects for given $L(t)$ of arbitrary time scale.

\subsubsection{Growing Domain}
Starting from the pattern state $A(\xi,t)=a(t)e^{iQ\xi}$, with $a\in\R$ and $Q\in\Z^+$, we can use \eqref{eq:RGLE_no_dilution} to obtain an evolution equation for the amplitude:
\begin{equation}\label{eq:RGLE_primary_nonlinear}
    \dot{a}  = \Tilde{\mu}(t)a - a^3,
\end{equation}
where 
\begin{equation}
\Tilde{\mu}(t) \equiv \mu - \frac{Q^2}{L(t)^2}.
\label{eq:mu_tilde}
\end{equation}
We restrict our attention to monotonically increasing $L(t)$. If $\mu > Q^2$, then the pattern state exists at $t=0$ and remains in existence for all $t$. For domain growth in which $L(t)\to\infty$ as $t\to\infty$, we see $a(t)\to\sqrt{\mu}$ regardless of the value of $Q$. If $\mu < Q^2$, then the trivial state $A=0$ is initially stable with respect to wave number $Q$ perturbations, but an increasing $L(t)$ can destabilize the trivial state at later times. To see this, we suppose $\mu$ is not too small, $L(t)$ increases sufficiently fast and $a(0)\ll 1$, so that the time for which $\Tilde{\mu}(t) \sim a^2$ is negligible. We may then linearize \eqref{eq:RGLE_primary_nonlinear} about the trivial state to obtain
\begin{equation}\label{eq:RGLE_primary_linearization}
    \dot{a} = \Tilde{\mu}(t)a,
\end{equation}
and so
\begin{equation}\label{eq:RGLE_primary_growing_delay}
    a(t) = a(0) \exp\left(\int_{0}^{t} \Tilde{\mu}(t')\,dt'\right).
\end{equation}
According to \eqref{eq:mu_tilde}, $\Tilde{\mu}(0) < 0$ and $\dot{\Tilde{\mu}}(t) > 0$ for all $t$. As a result, for a domain that grows sufficiently large, we define the \textit{turnaround time} $t_*>0$ such that $\Tilde{\mu}(t_*) = 0$. This denotes the time at which the system crosses the primary bifurcation. However, the system does not realize this bifurcation immediately. We define the \textit{exit time} $t_{exit} > t_*$ such that $a(t_{exit}) = a(0)$. This denotes the time when the perturbation exits its initial neighborhood. We can find $t_{exit}$ by solving
\begin{equation}\label{eq:entrance_exit}
    f(t_{exit}) \equiv \int_0^{t_{exit}} \Tilde{\mu}(t')\,dt' = 0,
\end{equation}
where $f(t)$ is the \textit{entrance-exit function}. We can also find the domain sizes at the corresponding times, denoted by $L_*$ and $L_{exit}$. Finally, we define
\begin{equation}
    t_{delay} \equiv t_{exit} - t_*
\end{equation}
to be the \textit{delay time}, or the extra time it takes for the perturbation to leave a neighborhood around the trivial state. We also define ${L_{delay}\equiv L_{exit}-L_*}$ to be the extra domain growth during the delay period. We refer to Fig.~\ref{fig:L_bifurcation_diagram} (inset) for a graphical depiction of this delay.

For an exponentially-growing domain $L(t) = e^{\sigma t}$, $\sigma > 0$, we obtain:
\begin{subequations}
\begin{align}
    t_* &= \frac{1}{\sigma}\ln\frac{Q}{\sqrt{\mu}}, \\
    L_* &= \frac{Q}{\sqrt{\mu}}, \\
    t_{exit} &= \frac{1}{2\sigma}\left[W_0\left(-\frac{Q^2}{\mu}e^{-Q^2/\mu}\right) + \frac{Q^2}{\mu}\right], \\
    L_{exit} &= \exp\left[\frac{1}{2}\left(W_0\left(-\frac{Q^2}{\mu}e^{-Q^2/\mu}\right) + \frac{Q^2}{\mu}\right)\right],\label{eq:L_exit} \\
    t_{delay} &= \frac{1}{2\sigma}\left[W_0\left(-\frac{Q^2}{\mu}e^{-Q^2/\mu}\right) + \frac{Q^2}{\mu} - \ln \frac{Q^2}{\mu}\right],\label{eq:t_delay} \\
    L_{delay} &= \exp\left\{\frac{1}{2}\left[W_0\left(-\frac{Q^2}{\mu}e^{-Q^2/\mu}\right)+\frac{Q^2}{\mu}\right]\right\} - \frac{Q}{\sqrt{\mu}},\label{eq:L_delay}
\end{align}
\end{subequations}
where $W_0(z)$ is the principal branch of the Lambert-W function with integral representation \cite{NIST}
\begin{equation}
    W_0(z) \equiv \frac{1}{\pi}\int_0^{\pi}\ln\left(1+z\frac{\sin t}{t}e^{t\cot t}\right)dt.
\end{equation}
A larger $\sigma$ results in a smaller $t_{delay}$ as shown in \eqref{eq:t_delay}. Additionally, $L_{exit}$ in \eqref{eq:L_exit} and $L_{delay}$ in \eqref{eq:L_delay} are independent of the growth rate $\sigma$.

In fact, the property that $L_{exit}$ and $L_{delay}$ are independent of the growth time scale holds for all types of monotonic domain growth. Applying \eqref{eq:entrance_exit}, we see that $L_{exit}$ solves
\begin{equation}\label{eq:entrance_exit_L}
    \int_{0}^{L_{exit}}\left(\mu - \frac{Q^2}{L^2}\right)\frac{1}{\dot{L}} \, dL = 0.
\end{equation}
For $\Tilde{L}(t) = L(\sigma t)$, $\sigma > 0$, we have $\dot{\Tilde{L}}(t) = \sigma\dot{L}(\sigma t)$, and \eqref{eq:entrance_exit_L} yields
\begin{align}
    &\int_{0}^{L_{exit}}\left(\mu - \frac{Q^2}{\Tilde{L}^2}\right)\frac{1}{\dot{\Tilde{L}}} \, d\Tilde{L}\nonumber \\
    = \frac{1}{\sigma}&\int_{0}^{L_{exit}}\left(\mu - \frac{Q^2}{L^2}\right)\frac{1}{\dot{L}} \, dL
    = 0.
\end{align}
Thus, $L_{exit}$ (and therefore, $L_{delay}$) is independent of the growth time scale.

\subsubsection{Shrinking Domain}

On a shrinking domain, we consider the problem in which the system begins in a pattern state where $\mu > Q^2$ but decays to the trivial state as $L(t)$ monotonically decreases. We are then interested in the \textit{entrance time} into a neighborhood around the trivial state; see the red line trajectory in Fig.~\ref{fig:L_bifurcation_diagram}. Specifically, we pick $\epsilon > 0$ and define $t_{enter}$ such that
\begin{equation}\label{eq:primary_shrinking_epsilon}
    a(t_{enter}) = \epsilon.
\end{equation}
Additionally, we redefine the delay time and corresponding domain delay as
\begin{subequations}
    \begin{align}
        t_{delay} &\equiv t_{enter} - t_*, \\
        L_{delay} &\equiv L_* - L_{enter}.
\end{align}
\end{subequations}
We interpret $L_{delay}$ as additional domain compression due to bifurcation delay.

\begin{figure}[!htbp]
\includegraphics[width=0.49\textwidth]{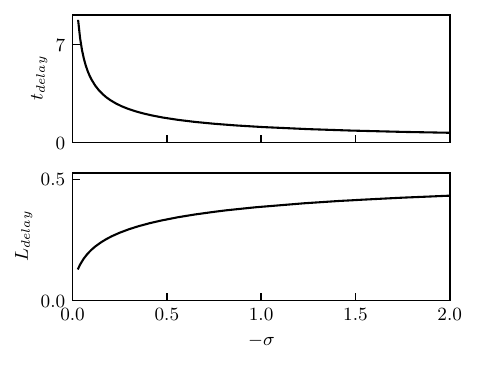}
\caption{On an exponentially shrinking domain ${L(t)=e^{\sigma t}}$ where $\sigma < 0$, as the domain shrinks faster, the delay time $t_{delay}$ decreases (top) but the length contraction $L_{delay}$ increases (bottom). The delay is calculated by comparing \eqref{eq:primary_shrinking_exponential} to the quasi-static case. Here, $\mu=3, \ Q=1$, and $\epsilon=0.0001$, where $\epsilon$ is the size of the neighborhood around the trivial state used for the delay calculation (see \eqref{eq:primary_shrinking_epsilon} and Fig.~\ref{fig:L_bifurcation_diagram}).}
\label{fig:primary_shrinking}
\end{figure}

Unlike the case of the growing domain, we cannot linearize the problem since the initial amplitude need not be small. Additionally, the amplitude, described by \eqref{eq:RGLE_primary_nonlinear}, departs from the time-independent pattern branch as the system approaches the primary bifurcation. Instead, as outlined in Appendix \ref{app:general_solution_primary_shrinking}, we obtain a general solution to \eqref{eq:RGLE_primary_nonlinear}. For the exponentially shrinking domain $L(t)=e^{\sigma t}$ with $\sigma < 0$, the closed form is
\begin{widetext}
\begin{equation}\label{eq:primary_shrinking_exponential}
    a(t) = \left[
    \frac{\exp\left(2\mu t + \frac{Q^2}{\sigma}\left(e^{-2\sigma t}-1\right)\right)}
    {-\frac{1}{\sigma}\exp\left(-\frac{Q^2}{\sigma}\right)\left(-\frac{\sigma}{Q^2}\right)^{-\mu/\sigma}\Gamma\left(-\frac{\mu}{\sigma}, -\frac{Q^2}{\sigma}, -\frac{Q^2}{\sigma}e^{-2\sigma t}\right) + a_0^{-2}}
    \right]^{1/2},
\end{equation}
\end{widetext}
where $\Gamma(s,t_0,t_1)$ is the generalized incomplete gamma function and $a_0=a(0)$.
Using this expression, we can understand how the rate of domain compression $\sigma$ affects the delay. We fix values for $\mu$, $Q$, and $\epsilon$ and compute the delay time $t_{delay}$ and associated domain compression $L_{delay}$ over a range of $\sigma$ by computing implicit solutions to \eqref{eq:primary_shrinking_exponential}. This is shown in Fig.~\ref{fig:primary_shrinking}. We see that as the domain shrinks faster, the delay time decreases while the domain compression increases.

\subsection{Secondary Bifurcation Delay: Eckhaus Instability and Phase Slip Generation}

\begin{center}
\begin{figure}
\includegraphics[width=0.49\textwidth]{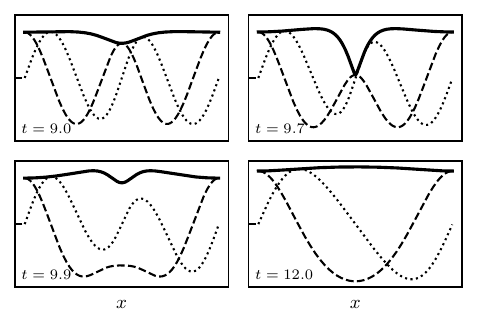}
\caption{An unstable $Q=2$ mode undergoes a phase slip leading to a stable $Q=1$ mode. The numerical simulation is run with $\Lambda=2\pi$, $L=1$, $\mu = 9$ and an initial perturbation $A'=0.01e^{ix}$ of the base state ${A=\sqrt{5}e^{2ix}}$.}
\label{fig:phase_slip}
\end{figure}
\end{center}

When a pattern state is perturbed, the system may transition into a more stable state via a phase slip that occurs when the pattern amplitude reaches zero at a point in the domain and the system deletes (or injects) a wavelength at that point. An example of this transition is depicted in Fig.~\ref{fig:phase_slip}.

For a given value of $\mu$, each wave number $Q$ pattern state is only unstable to modes $Q\pm k$ for select values of $k\in\Z^+$ and secondary bifurcations along the pattern branch stabilize the pattern state with respect to different values of $k$. On a fixed periodic domain, a perturbation of the pattern state by one of the unstable modes necessarily leads to a phase slip \cite{eckmann_phase_1995}, although its location depends on the initial perturbation profile. In nonperiodic systems, however, such as those with Robin boundary conditions, there is in general a preferred phase slip location determined by the shape of the dominant unstable mode \cite{ma_depinning_2012}.

On a time-dependent domain, the onset of phase slips can be delayed. Phase slips can also be prevented altogether---we call these \textit{arrested} phase slips. We aim to measure phase slip delay and classify the regimes in which phase slips are arrested.



\subsubsection{Fixed Domain}

To determine the stability properties of the pattern states, we let
\begin{equation}\label{eq:pattern_ptb}
    A(\xi,t) = A_Q(\xi) + A'(\xi,t),
\end{equation}
where
\begin{equation}\label{eq:pattern_base}
    A_Q(\xi) = \sqrt{\mu - \frac{Q^2}{L^2}}e^{iQ\xi}
\end{equation}
is the stationary wave number $Q$ pattern state and $A'(\xi,t)$ is a small perturbation. We write
\begin{equation}
    A'(\xi,t) = a_{k+}(t)e^{i(Q+k)\xi} + \overline{a}_{k-}(t)e^{i(Q-k)\xi},
    \label{eq:A_prime2a}
\end{equation}
where $k\in\Z^+$ and $a_{k+}(t),a_{k-}(t)\in\C$. Linearizing \eqref{eq:RGLE_no_dilution} about the base state $A_Q$ and inserting \eqref{eq:A_prime2a}, the evolution of this two-mode perturbation is given by
\begin{equation}
    \begin{pmatrix}
        {\dot{a}_{k+}} \\
        {\dot{a}_{k-}}
    \end{pmatrix}
    = {\bf{M}}\begin{pmatrix}
        {a_{k+}} \\
        {a_{k-}}
    \end{pmatrix},
\end{equation}
where
\begin{equation}
    {\bf{M}} = \begin{pmatrix}
        \mu - \frac{(Q+k)^2}{L^2} - 2a^2 & -a^2 \\
        -a^2 & \mu - \frac{(Q-k)^2}{L^2} - 2a^2
    \end{pmatrix}
\end{equation}
and $a=\sqrt{\mu-Q^2/L^2}$ is the amplitude of the base pattern state. The eigenvalues are
\begin{equation}
    \lambda_{k\pm} = -\left(\mu - \frac{Q^2}{L^2}\right) - \frac{k^2}{L^2} \pm \sqrt{\left(\mu-\frac{Q^2}{L^2}\right)^2 + \left(\frac{2Qk}{L^2}\right)^2}.
\end{equation}
The fast eigenvalue $\lambda_{k-}$ is always negative and governs the initial transient towards the eigenspace of the slow eigenvalue $\lambda_{k+}$. For $L=1$, the secondary bifurcations occur at $\mu=3Q^2-\frac{1}{2}k^2$ when $\lambda_{k+}=0$. These bifurcations are subcritical and generate unstable mixed mode states. The Eckhaus instability occurs at $k=1$ when the pattern state becomes stable with respect to all small perturbations of wave number $Q\pm k$.

\subsubsection{Shrinking Domain}
The secondary bifurcations stabilize unstable pattern states. A shrinking domain pushes a stable pattern state into the unstable regime. To a first approximation, we can study this transition via linearization about the pattern state as the perturbation amplitude is small near this transition. On the other hand, a growing domain requires the retention of the cubic nonlinearity because the deviation from the pattern state of interest is large. Thus, we first analyze the simpler case of the shrinking domain.

We recognize, however, that linearization may not be sufficient to fully characterize the delay on a shrinking domain even for small-amplitude perturbations. As described in {\cite{asch_slow_2023}} for a slowly-varying parameter, spatiotemporal resonances can excite modes more quickly than anticipated by linear analysis alone and these resonances can change the dominant mode of the perturbation and reduce the delay time of the Eckhaus instability.

Suppose the system starts in a stable wave number~$Q$ state with respect to $k$, i.e. ${\mu > 3Q^2 - \frac{1}{2}k^2}$. The equation that governs the evolution of the two-mode perturbation is now nonautonomous:
\begin{equation}\label{eq:nonautonomous}
    \begin{pmatrix}
        {\dot{a}_{k+}} \\
        {\dot{a}_{k-}}
    \end{pmatrix}
    = {\bf{M}}(t)\begin{pmatrix}
        {a_{k+}} \\
        {a_{k-}}
    \end{pmatrix},
\end{equation}
where
\begin{equation}
    {\bf{M}}(t) = \begin{pmatrix}
        \mu - \frac{(Q+k)^2}{L(t)^2} - 2a(t)^2 & -a(t)^2 \\
        -a(t)^2 & \mu - \frac{(Q-k)^2}{L(t)^2} - 2a(t)^2
    \end{pmatrix}
\end{equation}
and $a(t)$ is the time-dependent amplitude of the base pattern state that evolves by \eqref{eq:RGLE_primary_nonlinear}. The time-frozen eigenvalues of ${\bf{M}}(t)$ are given by

\begin{align}
    \lambda_{\pm}(t) =& \mu - \frac{Q^2}{L(t)^2} - 2a(t)^2 - \frac{k^2}{L(t)^2} \nonumber \\
    &\pm \sqrt{a(t)^4 + \left(\frac{2Qk}{L(t)^2}\right)^2}.
\end{align}
However, since the eigenspaces of $\lambda_{\pm}(t)$ are time-dependent, we cannot measure the bifurcation delay solely by examining the projection along the eigenspace of $\lambda_+(t)$ and the problem must be treated as the nonautonomous problem it is. In fact, there exist nonautonomous systems with exponentially growing solutions even when $\Re [\lambda_{\pm}(t)]<0$ for all $t>0$ {\cite{knobloch_enhancement_1992}}.

\begin{figure}
\centering
\includegraphics[width=0.49\textwidth]{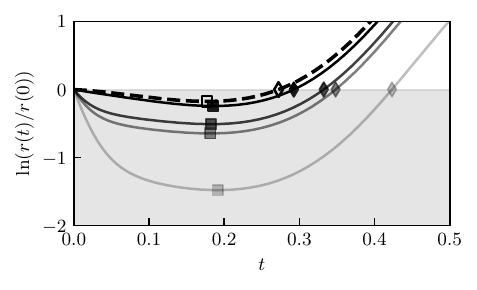}
\caption{The squared norm of different $k=1$ perturbations with $\arctan(|a_{k+}(0)/a_{k-}(0)|) = \{-\pi/4, 0, \pi/4, \pi/2\} $ (solid, decreasing opacity) of a ${Q=2}$ pattern state from simulations with $L(t)=e^{-2t}$ and $\mu=14$. The norms are bounded from above by {\eqref{eq:energy_method_bound} (dashed).}}
\label{fig:energy_method}
\end{figure}

To treat \eqref{eq:nonautonomous} as a nonautonomous problem, we apply an energy bound on the growth of a perturbation given by the maximum eigenvalue of ${\bf{M}}(t)+{\bf{M}^*}(t)$ \cite{farrell_generalized_1996}. Since ${\bf{M}}(t)$ is real and symmetric, we need only consider the least-stable time-frozen eigenvalue $\lambda_{+}(t)$. If we take the squared norm
\begin{equation}\label{eq:energy_method_norm}
    r(t) \equiv a_{k+}^2 + a_{k-}^2,
\end{equation}
then we can bound $r(t)$ by
\begin{equation}\label{eq:energy_method_bound}
    r(t) \leq r(0)\exp\left(2\int_{0}^{t}\lambda_{+}(\tau)\,d\tau\right).
\end{equation}
We can use this result to obtain a minimum delay time using the same techniques as outlined in the previous subsection. For some perturbations, the numerically computed $r(t)$ is very close to the upper bound \eqref{eq:energy_method_bound}. In other cases, $r(t)$ is much lower than the upper bound and results in a longer delay time as measured by the diamond symbols in Fig.~\ref{fig:energy_method}. However, the upper bound is respected by \emph{any} initial condition, as supported by the rapid contraction simulations shown in Fig.~\ref{fig:energy_method}.

\subsubsection{Growing Domain}

\begin{figure*}
\centering
\includegraphics[width=\textwidth]{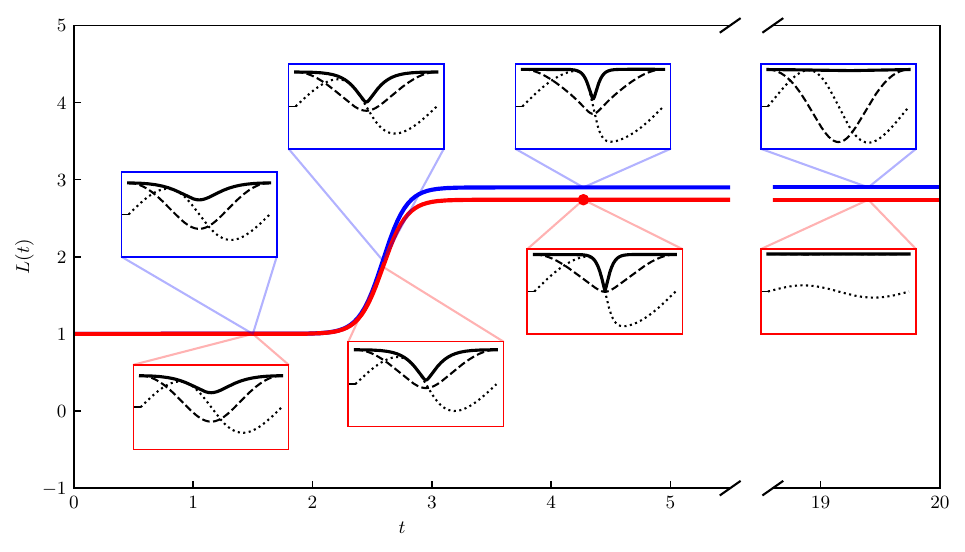} 
\caption{A phase slip is arrested with $L(t)=1.95+0.95\tanh(5(t-2.59))$ (blue, top) but develops with ${L(t)=1.85+0.85\tanh(5(t-2.59))}$ (red, bottom). The phase slip time is indicated by a solid red point. Both states are initialized with a $Q=1$ pattern state $A_Q(\xi)$ defined by \eqref{eq:pattern_base} and the same perturbation $A'=0.25e^{i\xi}$.}
\label{fig:arrested_phase_slip}
\centering
\end{figure*}

In the RGLE on a growing domain the passage across an Eckhaus bifurcation from the unstable to the stable side presents a challenge since we need to consider the basin of attraction of the pattern states. Figure~\ref{fig:arrested_phase_slip} demonstrates this in a dramatic way. For some $L(t)$, phase slips will develop even if the system crosses the Eckhaus instability. For other $L(t)$, phase slips are arrested---the system falls into the basin of attraction of the original pattern state. Thus, for a growing domain, we are not only concerned with the delay of a phase slip but also whether a phase slip develops at all.




Here, we outline a model for characterizing arrested phase slips on a time-dependent domain. For simplicity, we restrict our attention to the transition between the $Q=1$ pattern state and the $Q=0$ homogeneous state with $\Lambda=2\pi$. Without loss of generality, we can also force the (developing) phase slip to occur at $\xi=\Lambda/2=\pi$, i.e., in the domain center.

To physically motivate this model, we note that the Eckhaus instability is a phase instability, and the onset of a phase slip is characterized by a wave number diffusion equation with a negative diffusion coefficient \cite{cross_pattern_2009}. The amplitude is slaved to the phase; once the wave number compression is large enough, the amplitude reaches zero and a phase slip occurs. In the Eckhaus-stable regime, the mixed modes are the unstable states that resemble a developing phase slip (see the mixed mode profile in Fig.~\ref{fig:mu_bifurcation_diagram}). As $L(t)$ increases, the system becomes more Eckhaus-stable, and the mixed modes become more compressed. This suggests that we can use the wave number compression of the mixed modes to approximate the basin of attraction of an Eckhaus-stable pattern state. States which are more compressed than the relevant mixed mode fall outside the basin of attraction.

\begin{figure}
\includegraphics[width=0.49\textwidth]{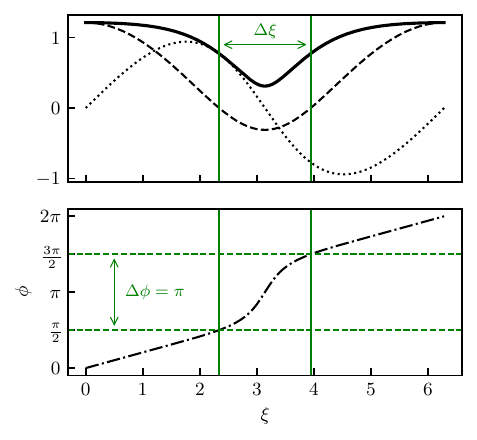}
\caption{The phase slip core width $\Delta\xi$ for a $Q=1$ profile, defined to be the spatial width around a phase slip in which there is an accumulation of $\Delta\phi=\pi$ phase. Top: $|A|$, $\Re A$, and $\Im A$. Bottom: phase $\phi$.}
\label{fig:core_width_visual}
\end{figure}

To measure the wave number compression at the location of the phase slip, we use the \textit{phase slip core width} $\Delta\xi$. To define this quantity, we write the profile in its amplitude-phase representation
\begin{equation}\label{eq:amplitude_phase}
    A(\xi,t) = a(\xi,t)e^{i\phi(\xi,t)}.
\end{equation}
For the phase of the $Q=1$ pattern, $\phi(\xi=0)=0$ and ${\phi(\xi=\Lambda)=2\pi}$. The phase slip occurs where $\phi=\pi$. Following the approach of \cite{tribelsky_phase-slip_1992}, the core width $\Delta\xi$ is defined to be the spatial distance between $\phi=\pi/2$ and $\phi=3\pi/2$, i.e. ${\Delta\phi=\pi}$ around the developing phase slip (see Fig.~\ref{fig:core_width_visual}). As $\Delta\xi$ decreases, the profile becomes more compressed. Once $\Delta\xi=0$, a phase slip occurs. On a fixed domain, $\Delta\xi$ scales as ${(t_{slip}-t)^{1/2}}$, where $t_{slip}$ is the time of the phase slip \cite{tribelsky_phase-slip_1992}.

\begin{center}
\begin{figure}
\includegraphics[width=0.49\textwidth]{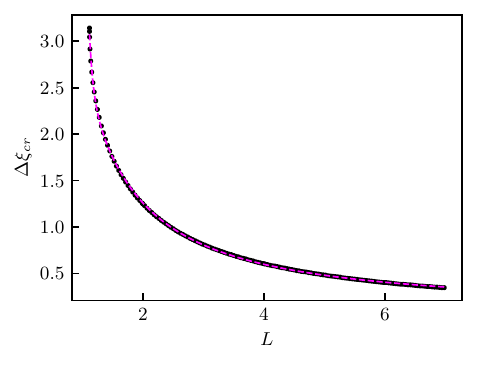}
\caption{The critical core width $\Delta\xi_{cr}$ of the $Q=1$ mixed mode as a function of $L$ when $\mu=2$, obtained by an interpolation of points obtained from numerical continuation in $L$.}
\label{fig:critical_core_width}
\end{figure}
\end{center}

To parametrize the basin of attraction, we compute the core width of each steady mixed mode at each $L$ for a fixed $\mu$, denoted as the \textit{critical core width} $\Delta\xi_{cr}(L)$. These are computed using pde2path, a Matlab package for numerical continuation and bifurcation analysis in systems of PDEs \cite{uecker_pde2path_2014, uecker2021numerical}. The spatial direction is discretized using the Fourier collocation method \citep{weideman2000matlab} following the implementation in \cite{uecker2021pde2path,liu2022staircase,liu2022single,liu2023fixed}, and we use $N=2048$ for the number of grid points. Figure~\ref{fig:critical_core_width} shows the interpolated pde2path results used to construct $\Delta\xi_{cr}(L)$ over $L$ for $\mu = 2$. As expected, as $L$ increases, $\Delta\xi_{cr}$ decreases. Thus, on a growing domain, the number of states that fall within the basin of attraction grows over time.

We construct the model as follows: given a perturbed $Q=1$ pattern state for some growing $L(t)$, we track the core width $\Delta\xi(t)$ over time. If a state falls within the basin of attraction at one time, then the state remains within the basin of attraction thereafter---this is an arrested phase slip. Specifically, 
\begin{itemize}
    \item if $\Delta\xi(t) > \Delta\xi_{cr}(L(t))$ at some time $t>t_*$, then the phase slip is arrested;
    \item if $\Delta\xi(t) < \Delta\xi_{cr}(L(t))$ for all time, then the phase slip develops.
\end{itemize}

\begin{figure*}
\centering
\includegraphics[width=\textwidth]{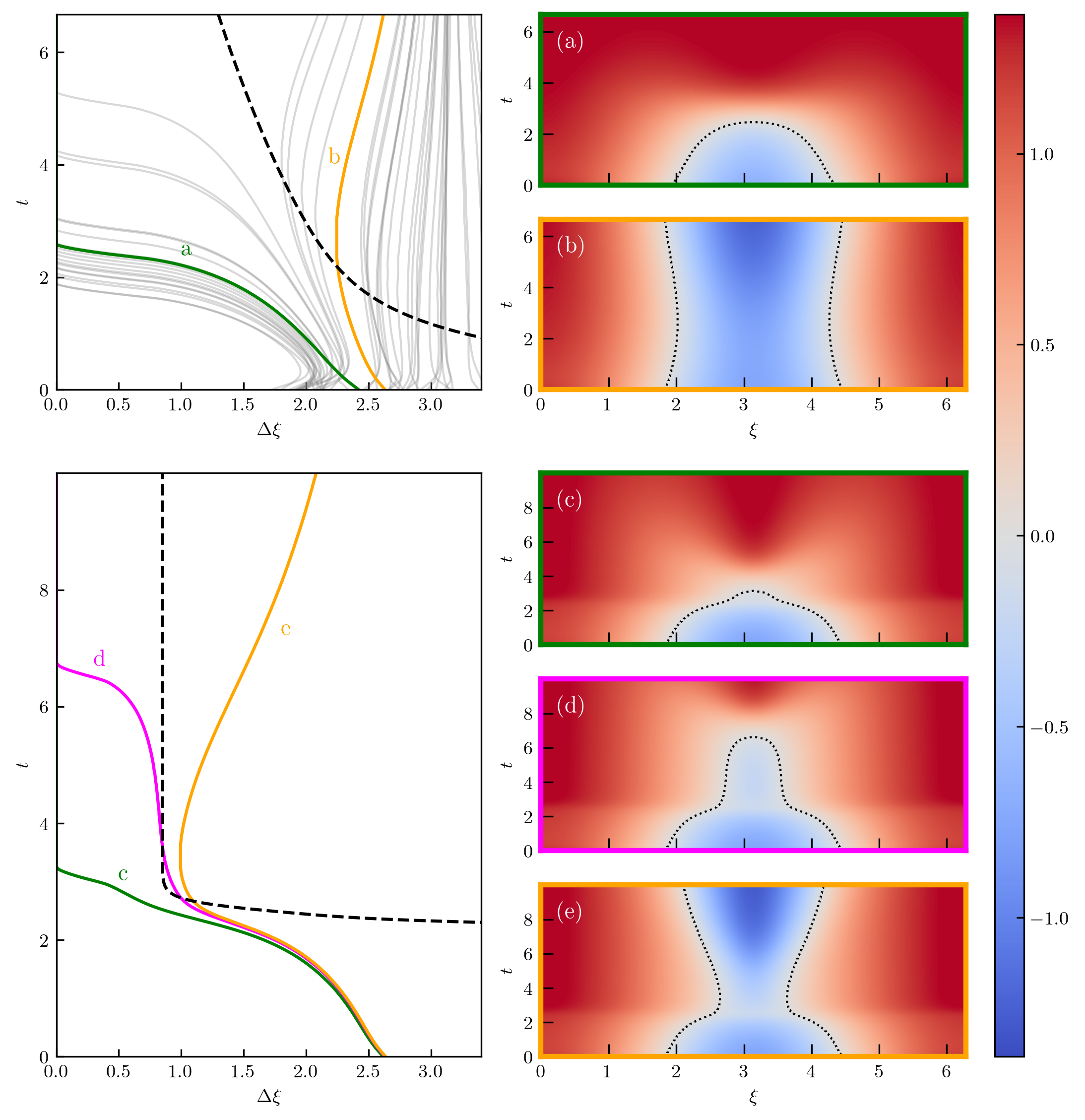} 
 \caption{The core width model classifies which regimes on a growing domain result in arrested phase slips between $Q=1$ and $Q=0$ by comparing the core width of the profile $\Delta\xi$ at each time to the critical core width $\Delta\xi_{cr}$ of the mixed mode. If $\Delta\xi>\Delta\xi_{cr}$ at some time, then the phase slip is arrested while $\Delta\xi=0$ indicates the presence of a phase slip. Left: the core width $\Delta\xi$ over time for various initial conditions (solid) and the critical core width $\Delta\xi_{cr}$ (dashed). Right: space-time plots of $\Re A$ for the selected profiles (colors), with the core width $\Delta\xi$ overlaid (dotted). Top: slowly-growing domain with $L(t)=e^{0.1t}$. The core width model works very well for 50 random finite-amplitude perturbations (gray lines). A phase slip (a) and arrested phase slip (b) are shown. Bottom: rapidly growing domain with $L(t)=1.93+0.93\tanh(5(t-2.59))$. Here the model is less successful. A phase slip (c) and arrested phase slip (e) are shown. The profile (d) reveals the limitations of this model, where $\Delta\xi>\Delta\xi_{cr}$ at some time but the phase slip occurs nonetheless.}
\label{fig:core_width_model}
\centering
\end{figure*}

We test this core width model with direct numerical simulations (DNS) of \eqref{eq:RGLE_no_dilution} using Dedalus, a Python library that uses spectral methods to solve differential equations \cite{burns_dedalus_2020}. We use a RK222 time-stepping scheme with a Fourier basis with $N=1024$. The linear terms are treated implicitly, but the time-dependent domain requires that the diffusion term is treated explicitly and a smaller time step is required for numerical stability. 

As shown in Fig.~\ref{fig:core_width_model}, we perform DNS with slow exponential growth $L(t)=e^{0.1t}$ and fast sigmoidal growth ${L(t)=1.93+0.93\tanh(5(t-2.59))}$. For exponential growth, the model exhibits remarkable accuracy for many randomly chosen perturbations. Evidently, the mixed mode core width acts as a useful criterion for determining whether phase slips are arrested.

However, examining Fig.~\ref{fig:core_width_model}(d), we find that this model is not perfect for fast sigmoidal domain growth, as a phase slip occurs even though the evolution (d) crosses $\Delta\xi_{cr}$. We find that the model is too lenient in classifying arrested phase slips because it does not account for transient effects due to changes in domain size. These transient effects cause additional compression beyond what is anticipated by the model. In practice, the range of perturbations for which this model fails is quite small, as demonstrated by the near-identical initial conditions shown in the bottom left panel of Fig.~\ref{fig:core_width_model}.

\section{Fronts}\label{sec:fronts}

Front propagation into unstable states arises from localized perturbations in spatially-extended systems. Fronts generally have a well-defined asymptotic velocity and profile. \textit{Pulled fronts} approach a spreading velocity determined by the linear dynamics ahead of the front, in contrast with pushed fronts that move faster than the linear spreading velocity due to nonlinear effects \cite{van_saarloos_front_2003}. The profile at the leading edge of the front is intimately connected to the selected velocity.

\begin{figure*}
\centering
\includegraphics[width=\textwidth]{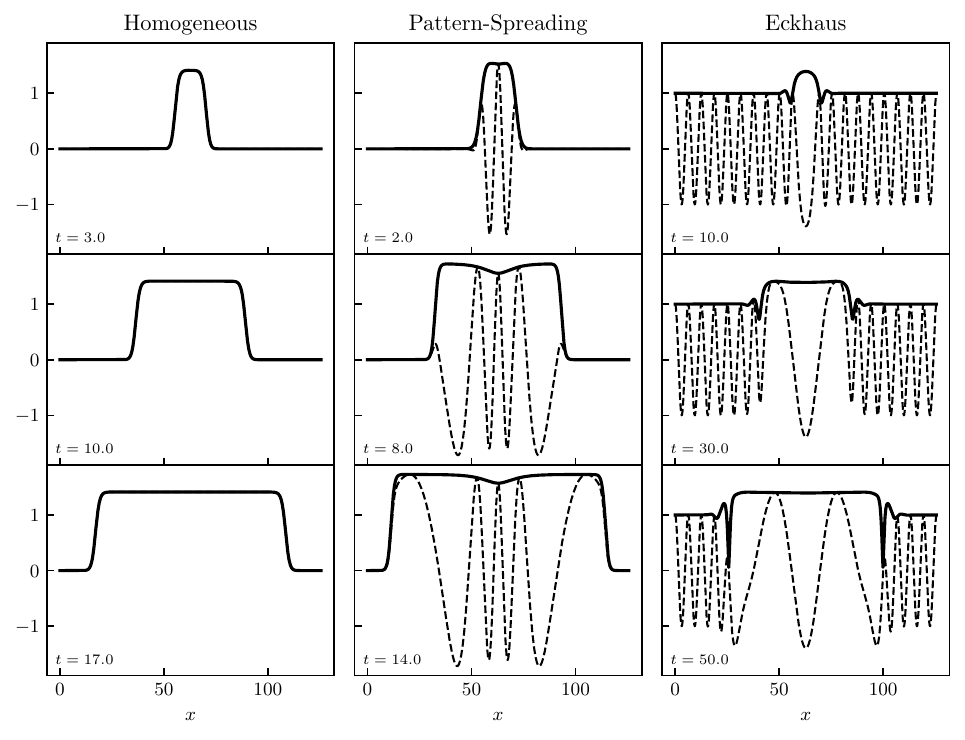} 
\caption{The time evolution of $|A|$ (solid) and $\Re A$ (dashed) of the different front types in the RGLE on a fixed domain $\Lambda=20\pi$. Values of $\mu$ are varied to emphasize qualitative features on the same spatial scale. Homogeneous fronts (left) are generated by the supercritical bifurcation at $\mu = 0$. Pattern-spreading fronts (middle) are generated by primary bifurcations with $Q\ne1$. Eckhaus fronts (right) are generated by secondary Eckhaus instability between two different pattern states. For visual clarity, the imaginary part is not shown.} 
\label{fig:fronts_overview}
\centering
\end{figure*}

In Fig.~\ref{fig:fronts_overview}, we identify three types of pulled fronts in the RGLE that correspond to following the three types of bifurcations:
\begin{enumerate}
    \item \textit{Homogeneous fronts} arise when the stable homogeneous state propagates into the unstable trivial state due to the supercritical bifurcation at $\mu=0$.
    \item \textit{Pattern-spreading fronts} arise when a pattern state propagates into the unstable trivial state due to the primary bifurcations at $\mu=Q^2$.
    \item \textit{Eckhaus fronts} arise when a stable pattern state propagates into an unstable pattern state due to the Eckhaus instability at $\mu=3Q^2-\frac{1}{2}$.
\end{enumerate}

Previous studies of pulled fronts in one dimension have mostly focused on their universal properties: the selected velocity and leading edge profile of pulled fronts are independent of the tracking height, specific nonlinearities, and initial conditions, provided the initial conditions are sufficiently steep \cite{van_saarloos_front_2003}. The asymptotic velocity is approached from below at a slow algebraic rate $O(t^{-1})$, while the asymptotic profile is approached at a rate $O(t^{-2})$ \cite{van_saarloos_front_2003}. The selected asymptotic front is the marginally stable uniformly propagating solution---this is known as the marginal stability criterion \cite{van_saarloos_front_2003}.

\begin{figure*}
\centering
\includegraphics[width=\textwidth]{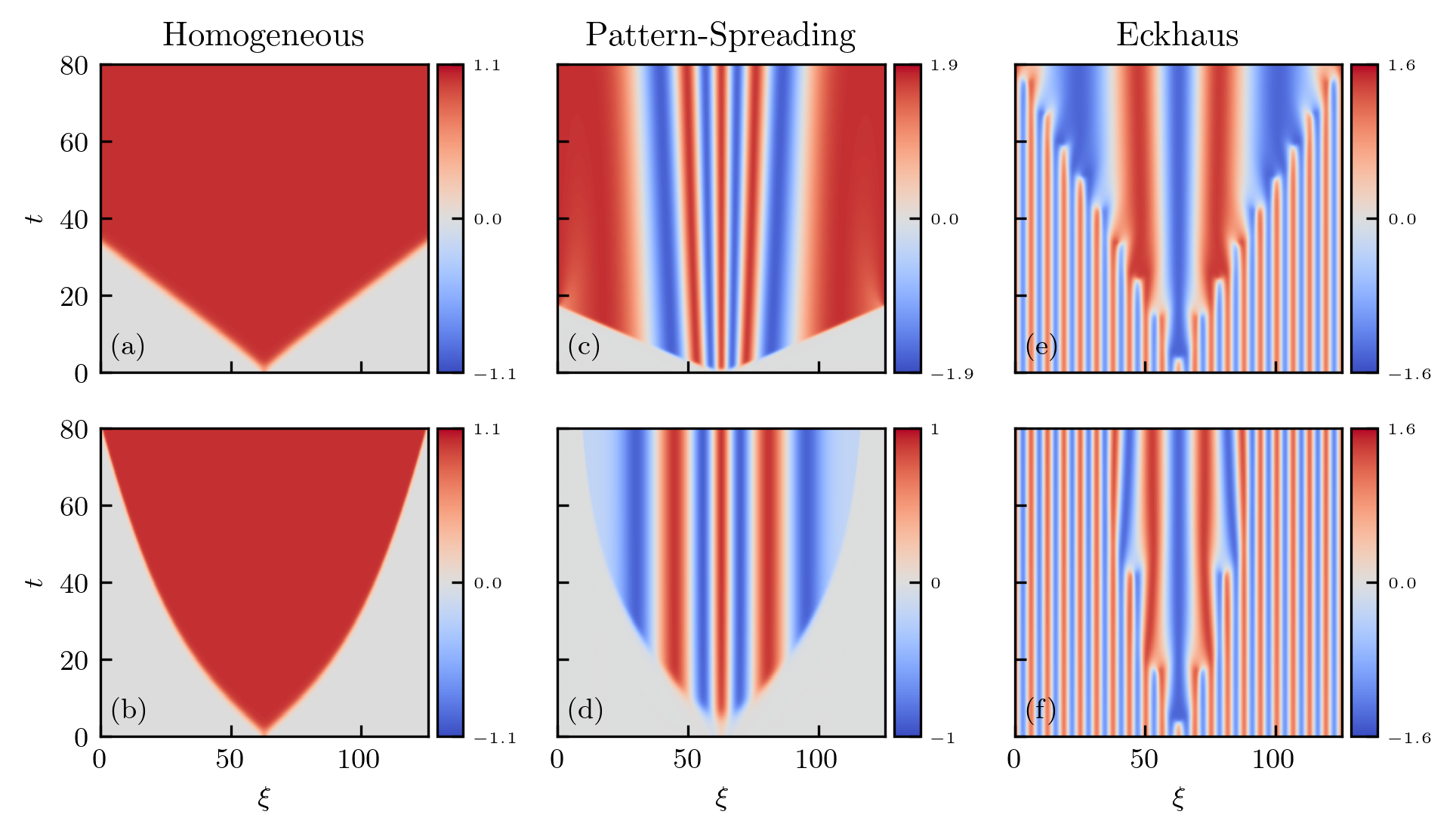} 
\caption{Space-time plots of $\Re A$ for homogeneous (left), pattern-spreading (middle), and Eckhaus (right) fronts on a fixed domain (top) and exponentially growing domain $L(t)=e^{\sigma t}$ (bottom) with $\Lambda=20\pi$. The values of $\mu$ and $\sigma$ are varied to emphasize qualitative features across the different types of fronts while retaining the same spatial and temporal scale. In all cases, the front slows down on the growing domain. For the homogeneous front, the initial condition is a narrow Gaussian. For the pattern-spreading front, the initial condition is a pattern modulated by a narrow Gaussian. The pattern is initially outside of the existence band at $t=0$ on the growing domain; therefore, front propagation is delayed. For the Eckhaus front, the initial condition is an Eckhaus-unstable pattern with a narrow Gaussian perturbation. On the growing domain the resulting phase slips occur less frequently and eventually stop, resulting in a phase-melting state.}
\label{fig:fronts_tdd}
\centering
\end{figure*}

However, on a time-dependent domain, it is unclear whether the asymptotic velocity and profile are meaningful at all---long-time asymptotic behavior cannot be described for general time-dependent domains. We cannot immediately determine the velocity and profile using the machinery of \cite{van_saarloos_front_2003}, which relies on long-time asymptotics. 
Therefore, we develop the theory from first principles to see which aspects of the fixed-domain theory must be adjusted. In Fig.~\ref{fig:fronts_tdd}, we give an overview of the homogeneous, pattern-spreading, and Eckhaus fronts in the RGLE on a time-dependent domain. In the subsections that follow, we analyze each front type in more detail. We primarily focus on homogeneous fronts because they are the most analytically tractable and most thoroughly analyzed in the literature \cite{van_saarloos_front_2003}. We make brief remarks about the effect of a growing domain on pattern-spreading and Eckhaus fronts at the end of this section.

\subsection{Homogeneous Fronts}\label{sec:fronts:homogeneous}

On a fixed domain, homogeneous or uniform fronts can be shifted into a comoving frame with time-independent behavior far from the front. If we take $A$ to be real, the RGLE is equivalent to the 1D Allen-Cahn equation or the Fisher-Kolmogorov-Petrovsky-Piskunov (F-KPP) equation with a cubic nonlinearity. Homogeneous fronts in these equations are the prototypical examples of pulled fronts and have been extensively studied on a fixed domain \cite{van_saarloos_front_2003, ebert_front_2000, aronson_multidimensional_1978}. However, on a time-dependent domain, we do not expect a stationary profile for any choice of a comoving frame and the profile will evolve over time.

\subsubsection{Linear Analysis}

Because the asymptotic velocity of a pulled front is entirely determined by the linear dynamics, we first analyze the linearization about the trivial state
\begin{equation}\label{eq:RGLE_linear}
    A_t = \mu A + \frac{1}{L(t)^2}A_{\xi\xi}.
\end{equation}
Without loss of generality, we take $A$ to be real. For simplicity, we perform this analysis on an infinite domain with a delta function initial condition and track the position of the point $A=C$. That is, we solve
\begin{equation}\label{eq:solve_for_C}
    C = A(\xi_C(t), t)
\end{equation}
for $\xi_C(t)$. There are two such points (see Fig.~\ref{fig:fronts_overview}); we focus on the right flank without loss of generality. From here, we can obtain the velocity $\dot{\xi}_C(t)$ and the profile $A(\xi_C(t), t)$. We characterize the profile by the \textit{steepness}, or spatial decay rate, of its leading exponential tail.

We first ignore time-dependent effects and obtain the asymptotic velocity and profile at each time from fixed-domain analysis. As shown in Appendix \ref{app:front_velocity}, the \textit{time-frozen asymptotic velocity} and \textit{steepness} of the front are
\begin{align}
    v^*(t) &= \frac{2\sqrt{\mu}}{L(t)}, \label{eq:linear_tdd_asym_vel} \\
    \lambda^*(t) &= \sqrt{\mu}L(t). \label{eq:linear_tdd_asym_steep}
\end{align}
These values may also be obtained from the marginal stability criterion \cite{van_saarloos_front_2003}. It is not immediately clear if these expressions remain meaningful in the presence of time dependence.

We may also solve \eqref{eq:RGLE_linear} directly using the dispersion relation $\omega(k,t)$ obtained by inserting ${A = \exp(ik\xi-i\int_0^t\omega(k,t') dt')}$ into \eqref{eq:RGLE_linear}:
\begin{equation}
    \omega(k,t) = i\left(\mu - \frac{k^2}{L(t)^2}\right).
\end{equation}
Thus, the time-evolved profile of an initial condition with Fourier transform $\tilde{A}(k,0)$ is given by:
\begin{widetext}
\begin{equation}
    A(\xi,t) = \frac{1}{2\pi}\int_{-\infty}^{\infty}\tilde{A}(k,0)\exp\left(ik\xi - i\int_0^t\omega(k,t') dt' \right) \,dk\,.
\end{equation}
\end{widetext}
For $A(\xi,0)=\delta(\xi)$ the exact solution is
\begin{equation}\label{eq:linear_tdd_xi}
    A(\xi,t) = \frac{1}{\sqrt{4\pi h(t)}}\exp\left(\mu t - \frac{\xi^2}{4h(t)}\right),
\end{equation}
where
\begin{equation}
    h(t) \equiv \int_0^t \frac{1}{L(t')^2} \,dt'.
\end{equation}

On a fixed domain, it is natural to move into a comoving frame with the asymptotic velocity \eqref{eq:linear_tdd_asym_vel} because this clears the growth term $\mu t$ in \eqref{eq:linear_tdd_xi} so that the amplitude neither grows nor decays exponentially. On a time-dependent domain, an immediate generalization is the time-frozen asymptotic frame
\begin{equation}
    \zeta = \xi - \int_0^t v^*(t') dt' = \xi - 2\sqrt{\mu}g(t),
\end{equation}
where
\begin{equation}
    g(t) \equiv \int_0^t \frac{1}{L(t')}\,dt'.
\end{equation}
However, the profile \eqref{eq:linear_tdd_xi} in this frame becomes
\begin{widetext}
\begin{equation}
    A(\zeta, t) = \frac{1}{\sqrt{4\pi h(t)}}\exp\left[-\frac{\zeta\sqrt{\mu}g(t)}{h(t)}-\frac{\zeta^2}{4h(t)}+\mu\left(t-\frac{g(t)^2}{h(t)}\right)\right],
\end{equation}
\end{widetext}
with a growth term that does not vanish because $g(t)^2 / h(t) \neq t$ except when $L$ is constant.


Instead, to clear the exponential growth in the comoving frame, we shift into the frame
\begin{equation}
    z = \xi - 2\sqrt{\mu t h(t)}\label{eq:z_frame}
\end{equation}
to obtain the profile
\begin{equation}
    A(z,t) = \frac{1}{\sqrt{4\pi h(t)}}\exp\left[ -z\sqrt{\frac{\mu t}{h(t)}} - \frac{z^2}{4h(t)} \right].\label{eq:linear_tdd_z}
\end{equation}
This is a much more natural frame---the exponential growth at $z=0$ vanishes, and the profile depends only on $h(t)$, the integrated diffusion coefficient, and not $g(t)$, the integrated square root of the diffusion coefficient.

Using \eqref{eq:z_frame} and \eqref{eq:linear_tdd_z}, we define the \textit{natural asymptotic velocity}
\begin{align}
    v^{**}(t) &= \frac{d}{dt}\left[2\sqrt{\mu t h(t)}\right] \nonumber \\
    & = \sqrt{\frac{\mu}{th(t)}}\left(h(t) + \frac{t}{L(t)^2}\right), \label{eq:linear_tdd_asym_vel2}
\end{align}
and the \textit{natural asymptotic steepness}
\begin{equation}
    \lambda^{**}(t) = \sqrt{\frac{\mu t}{h(t)}} \,.\label{eq:linear_tdd_asym_steep2}
\end{equation}
These are different from the time-frozen asymptotic values \eqref{eq:linear_tdd_asym_vel} and \eqref{eq:linear_tdd_asym_steep}. The natural asymptotic steepness $\lambda^{**}(t)$ is physically meaningful: it represents a balance between the growth $\mu t$ and diffusion $h(t)$. Note that $v^{**}=v^*$ and $\lambda^{**}=\lambda^*$ on a fixed domain.


We solve \eqref{eq:solve_for_C} using \eqref{eq:linear_tdd_z} to obtain the linear position and velocity
\begin{align}
    \xi_C(t) &= 2\left[h(t)\left(\mu t - \ln(C\sqrt{4\pi h(t)})\right)\right]^{1/2}, \label{eq:homogeneous_linear_position} \\
    \dot{\xi}_C(t) &= \frac{\frac{1}{L(t)^2}\left[\mu t - \ln(C\sqrt{4\pi h(t)})\right]+\mu h(t) - \frac{1}{2L(t)^2}}{\left[h(t)\left(\mu t - \ln\left(C\sqrt{4\pi h(t)}\right)\right)\right]^{1/2}}.\label{eq:homogeneous_linear_velocity}
\end{align}
Equations \eqref{eq:homogeneous_linear_position} and \eqref{eq:homogeneous_linear_velocity} are exact and valid for all times given a delta function initial condition. As expected, the velocity depends on the tracking point $C$, even at long times, because the changing steepness of the profile adds an amplitude-dependent ``rotation'' effect. On a constant domain, we recover the expected $O(t^{-1})$ approach to \eqref{eq:linear_tdd_asym_vel} from \eqref{eq:homogeneous_linear_velocity} as $t\to\infty$ for all $C$.

Finally, we compare these linear analysis results with the nonlinear results obtained from DNS. These DNS are run using a second-order central finite-difference scheme and RK4 time-marching scheme with ${\mu=1}$, ${\Lambda=100\pi}$, ${N=65536}$, and ${dt=10^{-5}}$, where $N$ is the number of grid points and $dt$ is the time step. This fine spatial discretization is necessary to precisely measure the nonlinear front velocity.

\begin{figure*}
\centering
\includegraphics[width=\textwidth]{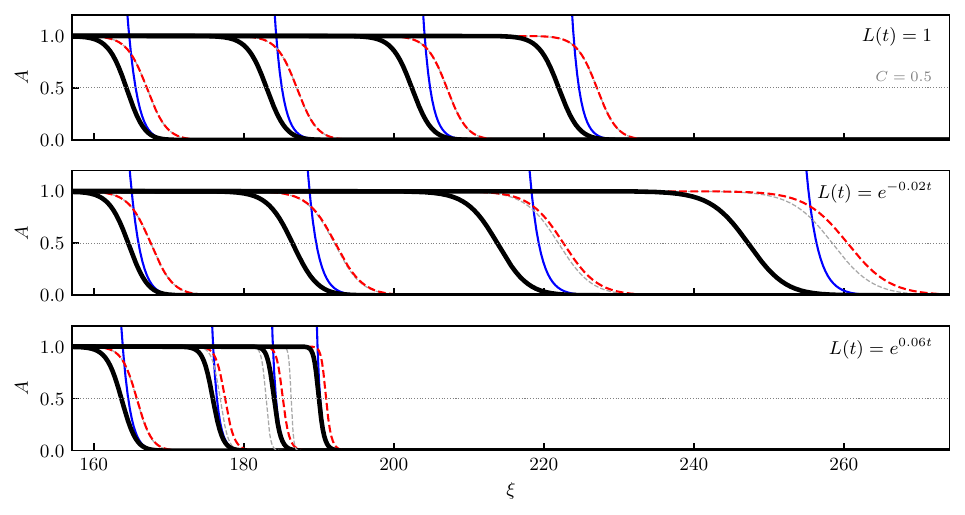}
  \caption{Homogeneous front propagation in the RGLE with $\mu=1$ at times $t=5, 15, 25, 35$ with various $L(t)$. Top: constant domain with $L(t)=1$. Middle: exponentially shrinking domain with $L(t)=e^{-0.02t}$. Bottom: exponentially growing domain with $L(t)=e^{0.06t}$. The black (solid, thick) lines are the nonlinear fronts computed with DNS. The blue (solid, medium) lines are the linear fronts calculated analytically from \eqref{eq:linear_tdd_xi}. The gray (dashed, thin) lines are the time-frozen asymptotic fronts. The red (dashed, medium) lines are the natural asymptotic fronts. In DNS, a delta function initial condition is approximated by a Gaussian $\frac{1}{\sqrt{2\pi s^2}}\exp\left[-(\xi-\Lambda/2)^2/2 s^2\right]$ with standard deviation $s=0.1$. This is narrow enough to model a delta function but wide enough to avoid numerical issues with finite differences. To allow fronts to propagate for long times without reaching the boundary, we select $\Lambda=100\pi$. This figure can be compared with Fig. 12 in \cite{van_saarloos_front_2003}.}
\label{fig:uniform_front_propagation}
\centering
\end{figure*}

\begin{figure*}
\centering
\includegraphics[width=\textwidth]{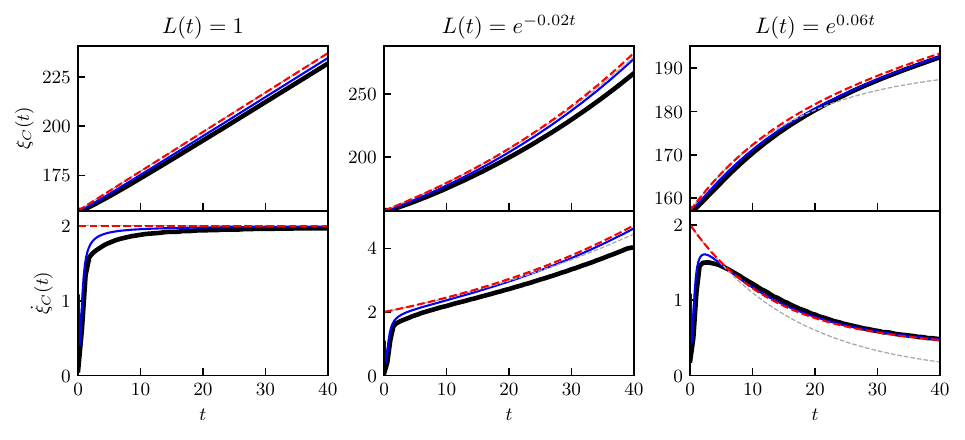}
 \caption{Velocities of the homogeneous fronts in Fig.~\ref{fig:uniform_front_propagation} with $C=0.5$. The black (solid, thick) lines denote the nonlinear position $\xi_C(t)$ and velocity $\dot{\xi}_C(t)$ obtained from DNS. The blue (solid, medium) lines denote the linear position \eqref{eq:homogeneous_linear_position} and velocity \eqref{eq:homogeneous_linear_velocity} calculated analytically. The gray (dashed, thin) lines denote the time-frozen asymptotic position and velocity \eqref{eq:linear_tdd_asym_vel}. The red (dashed, medium) lines denote the natural asymptotic position and velocity \eqref{eq:linear_tdd_asym_vel2}. This figure can be compared with Fig. 5 in \cite{ebert_front_2000}. On the fixed domain, both the linear and nonlinear velocities approach the constant $v^{**}=2$. In the shrinking domain, the nonlinear velocity deviates from the asymptotic and linear velocities as time increases. In the growing domain, the linear and nonlinear velocities overshoot the asymptotic velocities.}
\label{fig:uniform_front_velocities}
\centering
\end{figure*}

To explain the analysis shown in Figs. \ref{fig:uniform_front_propagation}, \ref{fig:uniform_front_velocities}, and \ref{fig:uniform_front_profiles}, we first summarize the definitions we use:
\begin{itemize}
    \item The \textit{nonlinear front} is obtained from DNS. The \textit{nonlinear velocity} and \textit{nonlinear profile} are the true, physical quantities---we use the other named velocities and profiles defined below for comparison.
    \item The \textit{linear front} is obtained by solving the linearized equation \eqref{eq:RGLE_linear} with a delta function initial condition. The \textit{linear velocity} \eqref{eq:homogeneous_linear_velocity} and \textit{linear profile} \eqref{eq:linear_tdd_z} are exact results.
    \item The \textit{time-frozen asymptotic front} is obtained from the general fixed-domain theory of fronts as described in Appendix \ref{app:front_velocity}. It is characterized by the \textit{time-frozen asymptotic velocity} $v^*(t)$ \eqref{eq:linear_tdd_asym_vel} and \textit{time-frozen asymptotic steepness} $\lambda^*(t)$ \eqref{eq:linear_tdd_asym_steep}. The steepness describes the leading edge of the full \textit{time-frozen asymptotic profile} which we compute below.
    \item The \textit{natural asymptotic front} is obtained by clearing the exponential growth term. It is characterized by the \textit{natural asymptotic velocity} $v^{**}(t)$ \eqref{eq:linear_tdd_asym_vel2} and \textit{natural asymptotic steepness} $\lambda^{**}(t)$ \eqref{eq:linear_tdd_asym_steep2}. We also compute the \textit{natural asymptotic profile} below. On a fixed domain, in which the time-frozen and natural fronts are the same, we use the general terms \textit{asymptotic front}, \textit{asymptotic velocity}, \textit{asymptotic steepness}, and \textit{asymptotic profile}.
\end{itemize}

In Fig.~\ref{fig:uniform_front_propagation}, we plot the propagating linear and nonlinear fronts in space for a fixed domain with ${L(t)=1}$, an exponentially growing domain with ${L(t)=e^{0.06t}}$, and an exponentially shrinking domain with ${L(t)=e^{-0.02t}}$. We also plot the time-frozen and natural asymptotic fronts; we compute these asymptotic profiles below. On the fixed domain, the linear front recedes from the asymptotic front. This is the expected logarithmic shift due to the $O(t^{-1})$ approach to the asymptotic velocity \cite{van_saarloos_front_2003}. The nonlinear front recedes more than the linear front; this can be derived via asymptotic matching \cite{ebert_front_2000}. For the shrinking domain, we observe a similar recession of the nonlinear front. However, in this case, the nonlinear front recedes faster as time goes on. We observe the opposite behavior on the growing domain.

In Fig.~\ref{fig:uniform_front_velocities}, we plot the linear, nonlinear, time-frozen asymptotic, and natural asymptotic front positions and velocities for these same regimes with tracking height $C=0.5$. On the constant domain, both the linear and nonlinear velocities approach the asymptotic velocity $v^{**}=2$. On the shrinking domain, the front velocities increase. At a basic level, this result is evident in the Eulerian (lab) frame: the front speed stays roughly constant while the distance between points decreases. Thus, it takes less time for the front to travel between two points. Likewise, on a growing domain, the distance between points increases in the Eulerian frame, so the front velocities decrease in the Lagrangian frame.

However, the detailed behavior of the nonlinear front velocity is rather complex. On the shrinking domain, the nonlinear front velocity departs from both the linear velocity \eqref{eq:homogeneous_linear_velocity} and the natural asymptotic velocity \eqref{eq:linear_tdd_asym_vel2} at large times. On the growing domain, the nonlinear front moves faster than the natural asymptotic front \eqref{eq:linear_tdd_asym_vel2} after around $t=5$. We would not expect any overshoot of the asymptotic velocity starting from steep initial conditions on a fixed domain: this is a new phenomenon. In addition, the linear and nonlinear velocities track the natural asymptotic velocity instead of the time-frozen asymptotic velocity. This confirms that the natural asymptotic frame is the appropriate comoving frame, suggesting that the marginal stability criterion breaks down on a time-dependent domain.


\subsubsection{Nonlinear Analysis}

The asymptotic velocity of a homogeneous front in the nonlinear equation \eqref{eq:RGLE_no_dilution} is the same as the asymptotic velocity in the linear equation \eqref{eq:RGLE_linear} because homogeneous fronts are pulled fronts. However, this does not mean that the nonlinear velocity matches the linear velocity \eqref{eq:homogeneous_linear_velocity}, since, on a time-dependent domain, the asymptotic velocities need not describe the actual velocities. In fact, we saw above that even in the fixed-domain case, the nonlinear front has a slower relaxation to the asymptotic velocity compared to the linear front.

To explain the nonlinear behavior, we consider the relationship between front velocity and profile. The key idea is the following: a time-dependent domain changes the front profile, and changes in the profile steepness at the leading edge cause changes in the front velocity.

On a fixed domain, if the initial conditions are steeper than the asymptotic steepness \eqref{eq:linear_tdd_asym_steep2}, then the asymptotic velocity will be approached from below. The velocity will never exceed $v^{**}$, and the steepness will never drop below $\lambda^{**}$. However, if the initial conditions are less steep, then the asymptotic velocity and steepness are not useful values. Instead, the front will propagate with a velocity $v > v^{**}$ and maintain its shallow exponential tail \cite{van_saarloos_front_2003}.

On a shrinking domain, $\lambda^{**}(t)$ is a decreasing function. Thus, because we start with a delta function initial condition, the front will always remain steeper than the asymptotic steepness, so the asymptotic values remain valid. On a growing domain, $\lambda^{**}(t)$ is an increasing function. Thus, the profile may be steep enough at $t=0$, but at a later time, the profile may be shallower than the asymptotic steepness. In this case, the asymptotic values are no longer valid, and the front can move faster than $v^{**}(t)$.

The above argument only holds for the linear velocity, and the linear and nonlinear profiles do not necessarily match. To confirm that $v^{**}(t)$ is also associated with the nonlinear profile, we move into the natural asymptotic frame given by \eqref{eq:z_frame} to obtain
\begin{equation}\label{eq:RGLE_z_frame}
    A_t = \mu A + \frac{1}{L(t)^2}A_{zz} - A^3 + v^{**}(t)A_z.
\end{equation}
For a fixed time $t$, we seek steady solutions to \eqref{eq:RGLE_z_frame}. This amounts to solving a boundary-value problem with boundary conditions of ${A(z\to -\infty)=\sqrt{\mu}}$ and ${A(z\to\infty)=0}$ with phase condition ${A(z=0)=C}$. We split \eqref{eq:RGLE_z_frame} into a first-order system by writing ${u=A}$ and ${w=A_z}$:
\begin{equation}\label{eq:asymptotic_nonlinear_system}
    \begin{cases}
        u_z = w \\
        w_z = L(t)^2\left[-v^{**}(t)w - (\mu u-u^3)\right].
    \end{cases}
\end{equation}
There are two fixed points at $(0,0)$ and $(\sqrt{\mu},0)$ which are stable and unstable, respectively. The front solution is the heteroclinic orbit connecting the two fixed points, since this is the only solution that satisfies the boundary conditions. These heteroclinic profiles propagating with the natural asymptotic speed are shown in Fig.~\ref{fig:uniform_front_propagation}. We can also obtain the time-frozen asymptotic profiles by moving into the frame associated with $v^*(t)$. The heteroclinic profiles in this frame are also shown in Fig.~\ref{fig:uniform_front_propagation}.

\begin{figure*}
\centering
\includegraphics[width=\textwidth]{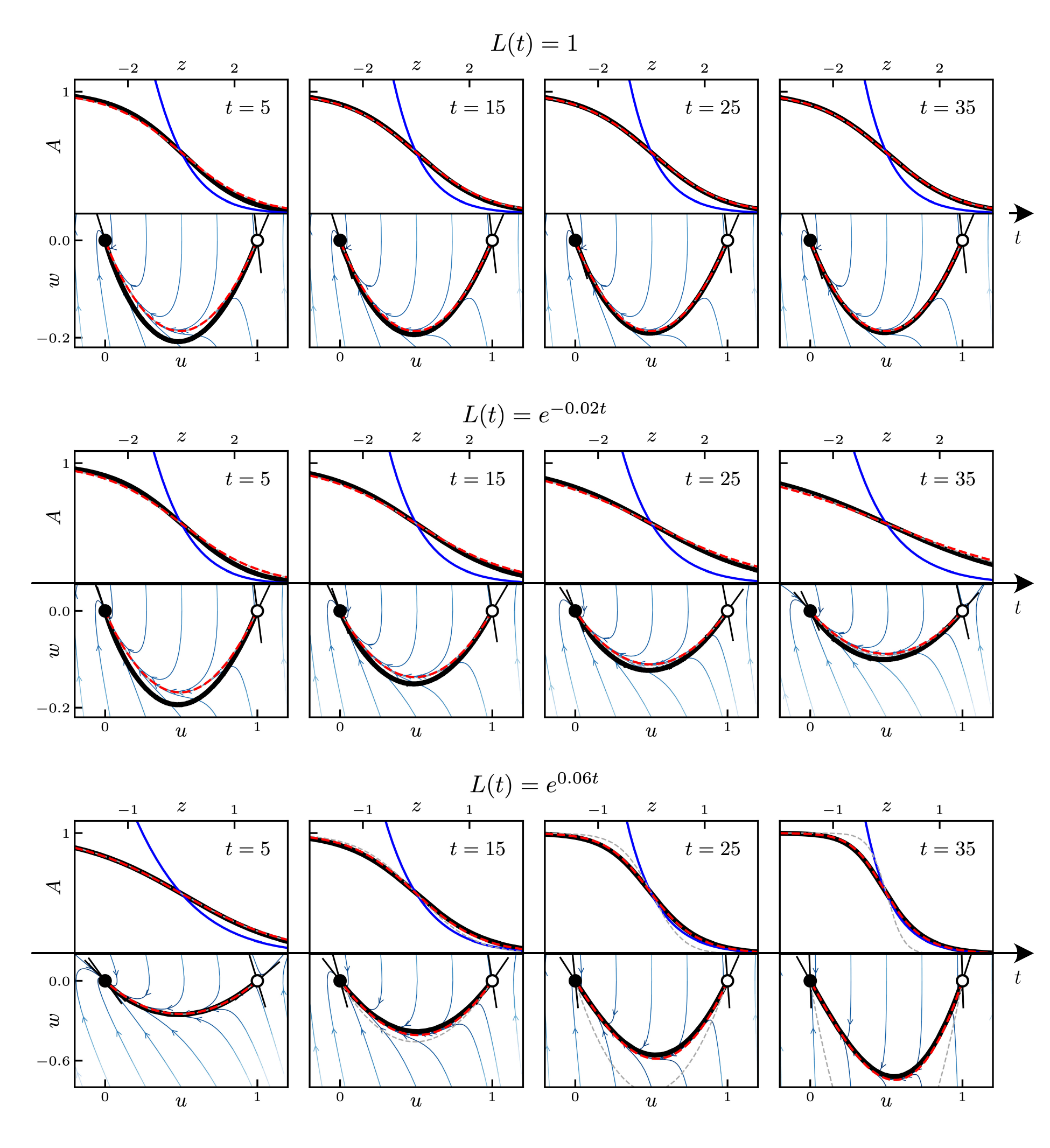}
 \caption{Profiles of the homogeneous fronts depicted in Fig.~\ref{fig:uniform_front_propagation}, where $\mu=1$. In each time snapshot, the top subplot displays overlapping nonlinear (black, solid, thick), linear (blue, solid, medium), time-frozen asymptotic (gray, dashed, thin), and natural asymptotic (red, dashed, medium) profiles intersecting at $C=0.5$. The bottom subplot shows the phase plane in the variables $(u,w)$ for stationary states in the natural asymptotic frame \eqref{eq:z_frame} with some trajectories plotted as described by \eqref{eq:asymptotic_nonlinear_system}. The two fixed points $(0,0)$ (stable, filled) and $(1,0)$ (unstable, empty) and their eigenspaces are shown. The fixed domain has a degenerate eigenspace for the $(0,0)$ fixed point which arises due to a double root. This is the well-known marginal stability property of the asymptotic nonlinear profile \cite{van_saarloos_front_2003}. However, on a time-dependent domain, $(0,0)$ loses its double root and becomes a generic stable node. In the growing domain, the time-frozen asymptotic profile, which tracks the degenerate state for all time, begins to deviate strongly from the other profiles. This is a further confirmation that $v^{**}(t)$ is the natural asymptotic velocity, not $v^{*}(t)$. Note that, for the growing domain, the scales of the plots are different. This figure can be compared with Fig.~4 in \cite{ebert_front_2000} and Fig.~9 in \cite{van_saarloos_front_2003}.}
\label{fig:uniform_front_profiles}
\centering
\end{figure*}
In Fig.~\ref{fig:uniform_front_profiles}, we overlap the linear, nonlinear, time-frozen asymptotic, and natural asymptotic profiles at $C=0.5$ at various times for the different $L(t)$. We also plot the phase plane for \eqref{eq:asymptotic_nonlinear_system} at each time with the two fixed points and their linearized eigenspaces. On top of this, we plot the trajectories of the nonlinear profile, time-frozen asymptotic profile, and natural asymptotic profile. From here, we can explain the nonlinear velocities shown in Fig.~\ref{fig:uniform_front_velocities}. On the shrinking domain, the nonlinear profile remains steeper than the natural asymptotic profile at all times which explains why the nonlinear velocity remains below the natural asymptotic velocity. On the growing domain, the nonlinear profile is less steep than the natural asymptotic profile at long times, so the nonlinear velocity is larger than the natural asymptotic velocity.

This analysis also demonstrates how the marginal stability criterion breaks down on a time-dependent domain. Note how, for the fixed domain, the linear profile does not line up closely with the nonlinear and natural asymptotic profiles. This is because, in this regime, the linearization at $(0,0)$ gives a double root with a degenerate eigenspace, so the asymptotic profile behaves like
\begin{equation}\label{eq:nonlinear_fixed_z_profile_large_z}
    A(z) \sim \ ze^{-\lambda^{**}z} \ (z\to\infty).
\end{equation}
This differs from the pure  $e^{-\lambda^{**}z}$ exponential tail from the linear analysis but is precisely what is prescribed by the marginal stability criterion. However, this degeneracy holds only in the fixed domain case. On a time-dependent domain, $(0,0)$ is a generic stable node in the natural asymptotic frame. It does not have a repeated spatial eigenvalue, so the natural asymptotic profile is not the marginally stable profile. 

\subsection{Pattern-spreading Fronts}

Pattern-forming fronts propagating into a trivial state are well-studied in a variety of systems \cite{van_saarloos_front_2003}. For the RGLE, velocity and stability analyses were conducted in \cite{ben-jacob_pattern_1985} for a fixed domain. In fact, these fronts are somewhat artificial in nature. Because wavelength injection cannot occur in the absence of phase slips, the initial condition prescribes the maximum number of wavelengths \cite{ben-jacob_pattern_1985}. If we restrict our attention to localized perturbations (specifically, perturbations with compact support), then these can only prescribe a finite number of initial wavelengths and thus a pattern-spreading front can only propagate for a finite time and distance. The homogeneous front eventually takes hold and the analysis of the previous subsection then applies.

Figure~\ref{fig:fronts_tdd} compares the resulting wave number evolution on a growing domain with that on a fixed domain. A growing domain has two qualitative effects on the front dynamics. First, the curved envelope in Fig.~\ref{fig:fronts_tdd}(d) indicates a decreasing velocity over time; this is very similar to the homogeneous front in Fig.~\ref{fig:fronts_tdd}(b). Second, by comparing the corresponding wavelengths between panels (c) and (d), we see that the local wavelength in the growing domain is larger than that on the fixed domain. This is unexpected since a slower spreading speed reduces wavelength stretching, leading to the expectation that, in the absence of phase slips, the observed wavelength will be smaller. Evidently, this effect competes with the preference for long wavelengths in this problem, which are permitted to develop in the presence of slower spreading speeds, and the growth of the domain gradually changes the competition between these two effects.

Owing to primary bifurcation delay, the propagation of the pattern-spreading front does not begin until the domain grows large enough for the local wave number to fall within the existence band. Thus, even though the rather simple delay analysis of \eqref{eq:RGLE_primary_nonlinear} does not appear to highlight any spatiotemporal features, it plays an important role in the timing of front propagation.

\subsection{Eckhaus Fronts}

As described in Section \ref{sec:bifurcation_delay}, an Eckhaus-unstable pattern state can undergo one or more phase slips to evolve into a stable pattern state with fewer wavelengths. If the initial perturbation is sufficiently localized, a propagating front of repeated phase slips can form (Fig.~\ref{fig:fronts_overview}). Some fronts of this type are constructed in \cite{eckmann_front_1993} but are not necessarily temporally stable or realized from localized initial conditions. Unlike the front types discussed so far, here the front invades an unstable state that is nontrivial and nonhomogeneous, although homogeneity can be recovered by transforming \eqref{eq:RGLE_no_dilution} into an amplitude-wave number representation. Remarkably, to high numerical accuracy, Eckhaus fronts travel at the linear spreading velocity even though phase slips are a nonlinear phenomenon \cite{dee_dynamical_1985}. 

In general, a moving front deposits a nonzero wave number. This wave number is selected dynamically and the resulting state may or may not be stable. In the present problem we saw that the homogeneous front deposits a zero wave number state, but this is no longer the case for Eckhaus fronts (Fig.~\ref{fig:fronts_overview}). Close to the Eckhaus boundary, wave number selection is described by the Cahn-Hilliard equation, and no phase slips take place \cite{gelens_coarsening_2010, hoyle_pattern_2006}. However, farther into the unstable regime, phase slips start to occur and determine the deposited wave number \cite{dee_dynamical_1985}. These phase slips occur either irregularly or periodically behind the moving front that continues to move with the speed predicted by the marginal stability prescription.
Since the dynamics of phase slips are modified on a time-dependent domain, as described in the first part of this paper, we may expect that time-dependence will likewise affect both the Cahn-Hilliard regime and the transition to and subsequent behavior of the phase slip regime. On a time-dependent domain the former is described by the phase equation (equivalently an equation for the wave number $k\equiv\phi_x$) derived in \cite{knobloch_stability_2014}, regularized by a fourth order linear term $k_{xxxx}$. While a study of this equation is beyond the scope of this paper, we focus here on the initial wave number selection process by an Eckhaus front on a time-dependent domain.

\begin{figure}
\centering
\includegraphics[width=0.49\textwidth]{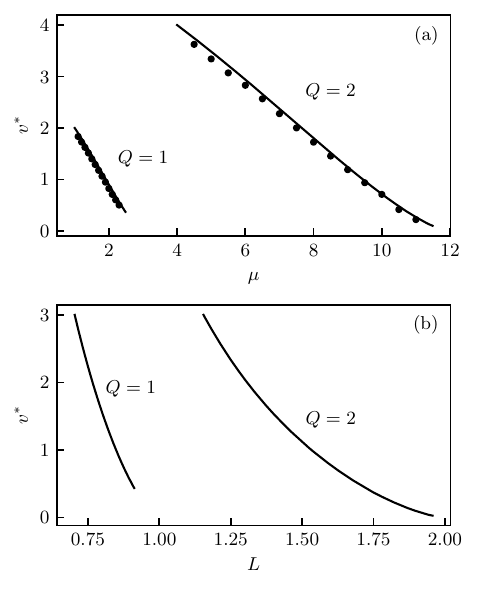} 
\caption{Eckhaus front velocities for $Q=1$ and $Q=2$ initial states. The solid lines are obtained from calculations in Appendix \ref{app:front_velocity_eckhaus}, and solid points denote measured velocities from DNS. In (a), the velocity is obtained for various $\mu$ with the constant domain $L=1$. In (b), the velocity is obtained for various values of $L$ with fixed $\mu=3$. This is a quasi-static analysis and on these small domains does not reveal any time-dependent behavior arising from domain growth.}
\label{fig:eckhaus_front_velocities}
\centering
\end{figure}

For this purpose we first look at the quasi-static analysis. In Fig.~\ref{fig:eckhaus_front_velocities}(a), we plot the linear spreading velocity as a function of $\mu$ for $L=1$ with initial wave numbers $Q=1$ and $Q=2$. The full derivation can be found in Appendix \ref{app:front_velocity_eckhaus}. We also add DNS data confirming the theoretical result. We see that for larger $\mu$, the Eckhaus front velocity is smaller. In Fig.~\ref{fig:eckhaus_front_velocities}(b), we plot the velocity as a function of $L$ for $\mu=3$. As expected, $L$ plays a similar stabilizing role as $\mu$: a larger domain size also results in a decrease in the front velocity.

Using DNS, we are able to extract some time-dependent behavior beyond the quasistatic regime. On an exponentially growing domain, we find that the Eckhaus fronts slow down as expected; compare (e) and (f) in Fig.~\ref{fig:fronts_tdd}. Once the domain grows to an Eckhaus-stable size, phase slips no longer occur and the front halts. The system then enters a \textit{phase-melting state} in which two different stable wave numbers temporarily coexist. The subsequent melting into a state with a uniform wave number is described in \cite{collet_solutions_1992}, but we do not yet know how these states evolve on a time-dependent domain. We also do not yet understand how the time-dependent Eckhaus front velocity may deviate from that shown in Fig.~\ref{fig:eckhaus_front_velocities}(b). Additionally, we expect delayed front propagation when crossing the Eckhaus instability but this topic is also beyond the scope of this paper.

\section{Dilution}\label{sec:dilution}

Dilution plays an important role in the changing stability of solutions on a time-dependent domain. Examining Figs.~\ref{fig:mu_bifurcation_diagram} and \ref{fig:L_bifurcation_diagram}, we see:
\begin{itemize}
    \item For a fixed $L$, increasing $\mu$ makes the trivial state less stable and the pattern states more stable.
    \item For a fixed $\mu$, increasing $L$ makes the trivial state less stable and the pattern states more stable.
\end{itemize}
Thus, $\mu$ and $L$ play similar roles in the stability of pattern solutions. If $L(t)$ is growing, then in the undiluted regime \eqref{eq:RGLE_no_dilution}, we expect the pattern states to become more stable. However, including dilution as in \eqref{eq:RGLE} changes the growth rate coefficient:
\begin{equation}
    \mu \mapsto \mu - \frac{\dot{L}(t)}{L(t)}.
\end{equation}
Thus dilution acts to decrease $\mu$ when the domain is growing, making the pattern states less stable. The reverse applies in the case of a shrinking domain. 
Thus dilution resists the changing stability of solutions due to time-dependence of the domain.

\begin{figure}
\centering
\includegraphics[width=0.49\textwidth]{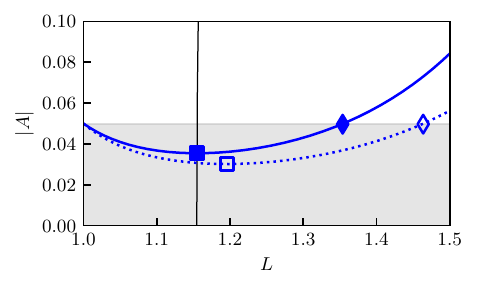} 
\caption{Dilution increases primary bifurcation delay. The system with dilution (dotted) has a larger turnaround size $L_*$ and exit size $L_{exit}$ than that without (solid). Both systems are initialized with the same initial condition and employ the same exponentially growing size, $L(t)=e^{0.02t}$, the same as the inset of Fig.~\ref{fig:L_bifurcation_diagram}.} 
\label{fig:dilution_bifurcation_delay}
\centering
\end{figure}

Consequently, we expect that dilution increases the delay time. We show this for a growing domain across a primary bifurcation. Suppose $\mu < Q^2$ and $\dot{L} > 0$. From the definition of the original turnaround time $t_*$, we have
\begin{equation}
    \mu - \frac{Q^2}{L(t_*)^2} = 0.
\end{equation}
With dilution present
\begin{equation}
    \mu - \frac{\dot{L}(t_*)}{L(t_*)} - \frac{Q^2}{L(t_*)^2} < 0
\end{equation}
since $\dot{L} > 0$. Thus, the turnaround time with dilution occurs later than that without dilution. Additionally, the exit time $t_{exit}$ is a root of the original entrance-exit function:
\begin{equation}
    f(t_{exit}) \equiv \int_{0}^{t_{exit}} \left(\mu - \frac{Q^2}{L(t')^2}\right)dt' = 0.
\end{equation}
The new entrance-exit function gives
\begin{equation}
    f_{dilut}(t_{exit}) = f(t_{exit}) - \int_{0}^{t_{exit}} \frac{\dot{L}(t')}{L(t')}\,dt' < 0.
\end{equation}
Thus, as expected, the exit time in the dilution regime also occurs later than that without dilution. Figure~\ref{fig:dilution_bifurcation_delay} compares the delay with dilution to that without dilution for the primary bifurcation, confirming this result. Furthermore, for any given $\mu$, if we take $L(t) = e^{\mu t}$, then the bifurcation never takes place since the source term is eliminated by the dilution effect.

\begin{figure}
\centering
\includegraphics[width=0.49\textwidth]{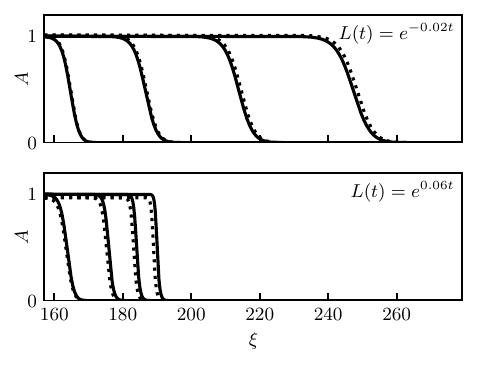} 
\caption{Propagating homogeneous fronts in the RGLE with dilution (dotted) and without dilution (solid) obtained from DNS. In the exponentially shrinking domain (top), the velocity increases faster when dilution is included. The reverse applies to the exponentially growing domain (bottom). The final amplitude behind the front also changes.} 
\label{fig:dilution_uniform_fronts}
\centering
\end{figure}

Dilution plays a different role for homogeneous fronts. If we modify the linear analysis in Eq.~\eqref{eq:linear_tdd_asym_vel2} to include the dilution term, we obtain the natural asymptotic velocity
\begin{equation}\label{eq:linear_tdd_asym_vel2_dilut}
    v_{dilut}^{**}(t) = 2\sqrt{M(t) h(t)},
\end{equation}
where
\begin{equation}
    M(t) \equiv \mu t - \int_{0}^{t} \frac{\dot{L}(t')}{L(t')}\,dt'.
\end{equation}
Recall that, in the undiluted regime, the natural asymptotic velocity $v^{**}(t)$ decreases on a growing domain and increases on a shrinking domain. With dilution, $M(t)$ is smaller for growing domains and larger for shrinking domains compared to its undiluted counterpart $\mu t$. Thus, dilution amplifies the effect of a time-dependent domain on the asymptotic speed. As demonstrated in Fig.~\ref{fig:dilution_uniform_fronts}, this result holds for exponential domains (constant $\dot{L}/L$) in the fully nonlinear regime.

Except for exponentially growing domains, extension of the nonlinear analysis to include dilution is not straightforward because the amplitude of the homogeneous solution becomes time-dependent. Thus, we can no longer seek stationary solutions in a comoving frame. Although we generally expect the dynamics behind a pulled front to play a minimal role in its velocity \cite{van_saarloos_front_2003}, a more comprehensive analysis is necessary to verify this claim for time-dependent domains.

\section{Discussion}\label{sec:discussion}

We extended previous work on bifurcation delay in the RGLE \cite{knobloch_stability_2014, knobloch_problems_2015} with analyses of the primary bifurcations for both growing and shrinking domains of length $L(t)$ with exact solutions and closed-form expressions. We also analyzed secondary bifurcation delay in detail. For the shrinking domain, we constructed an upper bound on the perturbation amplitude using an energy-based method to find a minimum delay time. For the growing domain, in which the nonlinear, infinite-dimensional dynamics must be retained, we outlined a heuristic model based on the core width of a phase slip to characterize the time-dependent basin of attraction of pattern states and validated this model with DNS. We used this model to classify arrested phase slips, which occur when the time scale of the domain growth competes with the Eckhaus instability.

We also gave a detailed linear analysis of homogeneous front propagation into an unstable trivial state on an arbitrary time-dependent domain starting from a delta-function initial condition. Our approach led to a new insight into what determines the front velocity. We defined the \textit{natural asymptotic velocity} and showed that in a time-dependent domain the velocity is no longer determined by the classical marginal stability criterion for pulled fronts, i.e. the degeneracy of the spatial eigenvalues of the trivial state breaks in the natural asymptotic frame. Our predictions for the nonlinear profile in this frame were corroborated by DNS of the RGLE. We also briefly examined DNS of pattern-spreading and Eckhaus fronts on an exponentially growing domain.

Lastly, we saw how dilution resists stability changes in a time-dependent domain, causing longer bifurcation delay compared to the undiluted regime. On the other hand, dilution amplifies the effect of a time-dependent domain on the velocity of homogeneous fronts.

Many aspects of bifurcation delay remain to be explored.  A more complete study and explanation of transient behavior would improve the accuracy of phase slip delays and arrests predicted in this paper. This is not well-explored because transients are unimportant in the fixed-domain RGLE: as long as perturbations are sufficiently small, all trajectories through a phase slip are determined by a one-dimensional eigenspace. Additionally, more complex domain time-dependence remains to be studied.  For example, oscillatory domain growth was not considered in this paper, although variants of the Ginzburg-Landau equation have been analyzed with a time-periodic parameter \cite{moehlis_eckhaus-benjamin-feir_1996}. Amplitude- and frequency-dependent shifts of the Eckhaus instability and mixed mode branches are expected, but this has not been considered in this paper. The effects of noise and imperfection terms are also of great interest \cite{benoit_dynamic_1991}.

Additional work remains to fully understand front dynamics in nonautonomous systems of this kind.  A larger body of numerical results is required for a complete catalog of the possible dynamics. Front velocities are notoriously difficult to measure in numerical simulations \cite{ponedel_front_2017, ebert_front_2000}. More precise and accurate numerical measurements of front velocities in a wider range of examples would verify our current understanding and perhaps identify new, unexpected behaviors. Theoretical progress is also needed to verify and explain the nonlinear front speeds and profiles obtained from DNS. Generalizing beyond a delta function initial condition on an infinite domain would enable better comparisons with the universal properties described in \cite{van_saarloos_front_2003}. Of course, we desire a reliable theory for the pattern-spreading and Eckhaus fronts as well. Developing a general theory for front propagation in nonautonomous partial differential equations would make a wide class of systems accessible to theory~\cite{rietkerk_evasion_2021, stoop_defect_2018}.

We are also interested in the effect of a time-dependent domain on the local dynamics of phase slips since these occur generically in growing patterns~\cite{ma_depinning_2012}. As described in Section~\ref{sec:bifurcation_delay}, the phase slip core width has an algebraic scaling law as the phase slip is approached \cite{tribelsky_phase-slip_1992}. It is unclear how this scaling behavior changes with domain growth.

This study illuminates countless possibilities for studying bifurcation delay and front propagation in more sophisticated models such as the complex Swift-Hohenberg equation \cite{gelens_coarsening_2010} and the Ginzburg-Landau equation with complex coefficients \cite{aranson_world_2002} on a time-dependent domain. These models include additional phenomena which have not been explored in detail on time-dependent domains, such as coarsening, spatially localized structures, traveling waves, and naturally occurring pattern-forming fronts. We also look towards bridging the gap between these models and observed phenomena in physical and biological realizations of patterns on time-dependent domains \cite{knobloch_problems_2015}. Quantitative predictions of bifurcation delay times and front propagation speeds in experiments would demonstrate the efficacy of this theory.

\begin{acknowledgments}
We gratefully acknowledge helpful comments and suggestions from V. Klika and an anonymous referee.
This work was supported by the 2022 UC Berkeley Physics Innovators Initiative (Pi$^2$) Scholars Program (T.T.), the Berkeley Physics-and-Astronomy Undergraduate Research Scholars (BPURS) Program (T.T.) and by the National Science Foundation under grants DMS-1908891 (B.F. and E.K.) and OCE-2023541 (C.L. and E.K.). C.L. acknowledges a discussion of pde2path with Tobias Frohoff-H\"{u}lsmann and support from the NASA CT Space Grant P-2104 and Connecticut Sea Grant PD-23-07 during the completion of this work.
\end{acknowledgments}


\appendix

\section{RGLE with a Conservation Law}\label{app:dilution}

The 1D real Ginzburg-Landau equation on a fixed domain is 
\begin{equation}
    A_t = \mu A + A_{xx} - |A|^2A,
\end{equation}
where $A$ is a complex variable and $x\in [0, \Lambda]$. 

If $A$ is a conserved quantity, we cannot use a regular time derivative when switching to a time-dependent domain $\Omega_t$. Instead, by the Reynolds transport theorem in one spatial dimension, we have
\begin{equation}\label{eq:reynolds}
    \frac{d}{dt}\int_{\Omega_t} A \, dV = \int_{\Omega_t} (A_t + uA_x + u_xA)\, dV,
\end{equation}
where $u$ is some velocity determined by the growing domain. This takes into account the fact that material elements change size.

We now restrict to isotropic growth in which ${u = \frac{\dot{L}}{L}x}$. Using the modified time derivative found in \eqref{eq:reynolds}, the full time-dependent RGLE becomes
\begin{equation}
    A_t + \underbrace{\frac{\dot{L}(t)}{L(t)}xA_x}_{\text{advection}} + \underbrace{\frac{\dot{L}(t)}{L(t)}A}_{\text{dilution}} = \mu A + A_{xx} - |A|^2A,
    \label{eq:RGLE_appendix_A}
\end{equation}
where $x\in [0, \Lambda L(t)]$. The second term represents advection and the third term represents dilution. 

Note that \eqref{eq:RGLE_appendix_A} describes the RGLE in the Eulerian (lab) frame, where the first two terms together equal the material derivative of $A$. Thus, when we change to the Lagrangian frame, we expect the material derivative to become a normal time derivative because the effect of advection is built into the Lagrangian frame. To show this, let
\begin{equation}
    A(x,t) = \Tilde{A}(\xi(x,t), t),
\end{equation}
where $\Tilde{A}$ is the amplitude in the Lagrangian frame and $\xi \in [0,\Lambda]$ is the Lagrangian coordinate, i.e. $\xi(x,t) = \frac{x}{L(t)}$. Then,
\begin{subequations}
\begin{align}
    A_t &= \Tilde{A}_t + \Tilde{A}_\xi \frac{d\xi}{dt} \\
     &= \Tilde{A}_t + \Tilde{A}_{\xi}\frac{d}{dt}\left(\frac{x}{L(t)}\right) \\
     &= \Tilde{A}_t + \Tilde{A}_{\xi}\left(-\frac{\dot{L}(t)}{L(t)^2}x\right) \\
     &= \Tilde{A}_t - \frac{\dot{L}(t)}{L(t)}\xi\Tilde{A}_{\xi}. 
\end{align}
\end{subequations}
Next, the advection term becomes
\begin{subequations}
\begin{align}
    \frac{\dot{L}(t)}{L(t)}xA_x &= \frac{\dot{L}(t)}{L(t)}\frac{x}{L(t)}(L(t)A_x) \\
    &= \frac{\dot{L}(t)}{L(t)}\xi\Tilde{A}_{\xi}.
\end{align}
\end{subequations}
Thus, as expected, the advection term drops out, and the Lagrangian description of the RGLE becomes
\begin{equation}
    \Tilde{A}_t + \frac{\dot{L}(t)}{L(t)}\Tilde{A} = \mu \Tilde{A} + \frac{1}{L(t)^2}\Tilde{A}_{\xi\xi} - |\Tilde{A}|^2\Tilde{A}.
\end{equation}
We drop the tildes to obtain \eqref{eq:RGLE}.

\section{Pattern Amplitude on a Shrinking Domain}\label{app:general_solution_primary_shrinking}

Here, we find the general solution to
\begin{equation}
    \dot{a} = \Tilde{\mu}(t)a - a^3,
\end{equation}
where $a \geq 0$. This is a Bernoulli-type equation which can be solved using the substitution $v=a^{-2}$ to obtain the explicit solution
\begin{equation}
    a(t) = \left[
    \frac{\exp\left(2 \int_0^t \Tilde{\mu}(t')dt'\right)}
    {2\int_0^t \exp\left(2 \int_0^{t'} \Tilde{\mu}(t'')dt''\right)dt' + a_0^{-2}}
    \right]^{1/2},
    \label{eq:a_t_shrinking_general}
\end{equation}
where $a(0)=a_0$. 

For an exponentially shrinking domain with $L(t)=e^{\sigma t}$, $\sigma < 0$, 
\begin{equation}
    \Tilde{\mu}(t) = \mu - \frac{Q^2}{L(t)^2} = \mu - Q^2e^{-2\sigma t}.
\end{equation}
To find the integral in the denominator, we observe
\begin{align}
    &\int_0^t \exp\left(2 \int_0^{t'} \Tilde{\mu}(t'')\,dt''\right)dt'\nonumber
    \\
    =\,& e^{-Q^2/\sigma}\int_0^t e^{2\mu t'}e^{(Q^2/\sigma)e^{2\sigma t'}} dt'.
\end{align}
Now let $a=2\mu$, $b=-Q^2/\sigma$, and $c=-2\sigma$. With the substitution $v=be^{ct'}$, we obtain
\begin{equation}
    \int_0^t e^{at'}e^{-be^{-ct'}} dt' = \frac{1}{c}\left(\frac{1}{b}\right)^{a/c}\Gamma\left(\frac{a}{c},b,be^{ct}\right),
\end{equation}
where
\begin{equation}
    \Gamma(s,t_0,t_1) = \int_{t_0}^{t_1} t^{s-1}e^{-t}\,dt
\end{equation}
is the incomplete generalized gamma function. Substituting for $a$, $b$, and $c$, we obtain
\begin{widetext}
\begin{equation}
    \int_0^t \exp\left(2 \int_0^{t'} \Tilde{\mu}(t'')dt''\right)dt' = -\frac{1}{2\sigma}\left(-\frac{\sigma}{Q^2}\right)^{-\mu/\sigma}\Gamma\left(-\frac{\mu}{\sigma}, -\frac{Q^2}{\sigma}, -\frac{Q^2}{\sigma}e^{-2\sigma t}\right).
    \label{eq:double_exp_integral}
\end{equation}
\end{widetext}
With this result Eq.~\eqref{eq:a_t_shrinking_general} gives \eqref{eq:primary_shrinking_exponential}.

\section{Homogeneous Front Velocity}\label{app:front_velocity}

We derive here the asymptotic velocity and steepness of a homogeneous front on a domain with fixed length $L$. There are many ways to do this; here, we use the general technique outlined in \cite{van_saarloos_front_2003}. First, we obtain the dispersion relation
\begin{equation}\label{eq:dispersion}
    \omega(k) = i\left(\mu - \frac{k^2}{L^2}\right).
\end{equation}
Now suppose we move into a comoving frame with velocity $v^*$. Our goal is to determine which value for $v^*$ allows the amplitude to neither grow nor decay. We determine this velocity by considering a complex wave number $k^*$ such that
\begin{equation}\label{eq:saddle_point}
    \dv{(\omega(k)-v^* k)}{k}\eval_{k^*} = 0.
\end{equation}
This $k^*$ is a saddle point in the complex plane that, in the long-time limit, provides the dominant contribution to the inverse Fourier transform needed to determine the physical amplitude. We also require that in this frame the amplitude neither grows nor decays:
\begin{equation}\label{eq:fixed_amplitude_front}
    \Im \omega(k^*) - v^*\Im k^* = 0.
\end{equation}
Putting \eqref{eq:saddle_point} and \eqref{eq:fixed_amplitude_front} together, we get
\begin{equation}\label{eq:marginal_stability_criterion}
    v^* = \dv{\omega(k)}{k}\eval_{k^*} = \frac{\Im\omega(k^*)}{\Im k^*}.
\end{equation}
Let $k^* = k_r^* + ik_i^*$, where $k_r, k_i \in \R$. Then, substituting \eqref{eq:dispersion} into \eqref{eq:marginal_stability_criterion}, we obtain
\begin{equation}\label{eq:marginal_stability_plugged_in}
    v^* = \frac{2k_i^*}{L^2} - \frac{2k_r^*}{L^2}i = \frac{1}{k_i^*}\left(\mu - \frac{(k_r^*)^2 - (k_i^*)^2}{L^2}\right).
\end{equation}
Separating into real and imaginary parts, we find that $k^* = i\sqrt{\mu}L$, and hence that
\begin{equation}
    v^* = \frac{2\sqrt{\mu}}{L}.
\end{equation}
The front steepness is given by
\begin{equation}
    \lambda^* = \Im k^* = \sqrt{\mu}L.
\end{equation}

\section{Eckhaus Front Velocity}\label{app:front_velocity_eckhaus}

We now derive the linear spreading velocity for Eckhaus fronts on a domain with fixed length $L$ \cite{dee_dynamical_1985}. Since these fronts propagate into an Eckhaus-unstable pattern, it is natural to write the amplitude-phase representation of $A$ as
\begin{equation}
    A(\xi,t) = a(\xi,t)e^{i\phi(\xi,t)}.
\end{equation}
Then, defining the wave number $q(\xi,t)\equiv\phi_{\xi}(\xi,t)$, we can rewrite the RGLE as
\begin{subequations}
\begin{align}
    a_t &= \left(\mu-\frac{q^2}{L^2}\right)a + \frac{1}{L^2}a_{\xi\xi} - a^3 \\
    q_t &= \frac{1}{L^2}\pdv{\xi}\left(q_{\xi} + 2q\pdv{\ln{a}}{\xi}\right).
\end{align}
\end{subequations}
The initial pattern state with uniform wave number $q_0=Q$ and amplitude $a_0=\sqrt{\mu-Q^2/L^2}$ corresponds to a fixed point of the $\xi$-independent equations.

The derivation of the Eckhaus front linear spreading velocity now follows similarly to the homogeneous case. Consider $a=a_0+a'$ and ${q=q_0+q'}$. After linearizing, we use the relations ${a'=a_1(k)e^{ik\xi-i\omega t}}$ and ${q'=q_1(k)e^{ik\xi-i\omega t}}$ to find the dispersion relation:
\begin{equation}
\begin{pmatrix}
    -2a_0^2-\frac{k^2}{L^2}+i\omega & -\frac{2Qa_0}{L^2} \\
    -\frac{2Qk^2}{a_0L^2} & -\frac{k^2}{L^2}+i\omega
\end{pmatrix}\begin{pmatrix}
    a_1(k) \\
    q_1(k)
\end{pmatrix} = 0.
\end{equation}
There are two branches of this relation,
\begin{equation}
    \omega_{\pm}(k) = i\left[-a_0^2-\frac{k^2}{L^2} \pm \sqrt{a_0^4+\frac{4Q^2k^2}{L^4}}\right].
\end{equation}
Taking the positive root $\omega_+(k)$, we seek a velocity where
\begin{equation}
    v^* = \dv{\omega_+(k)}{k}\eval_{k^*} = \frac{\Im\omega_+(k^*)}{\Im k^*}.
\end{equation}
We can solve this equation implicitly for a fixed $Q$ while varying either $\mu$ or $L$ to obtain Fig.~\ref{fig:eckhaus_front_velocities}.


\begin{thebibliography}{85}%
\makeatletter
\providecommand \@ifxundefined [1]{%
 \@ifx{#1\undefined}
}%
\providecommand \@ifnum [1]{%
 \ifnum #1\expandafter \@firstoftwo
 \else \expandafter \@secondoftwo
 \fi
}%
\providecommand \@ifx [1]{%
 \ifx #1\expandafter \@firstoftwo
 \else \expandafter \@secondoftwo
 \fi
}%
\providecommand \natexlab [1]{#1}%
\providecommand \enquote  [1]{``#1''}%
\providecommand \bibnamefont  [1]{#1}%
\providecommand \bibfnamefont [1]{#1}%
\providecommand \citenamefont [1]{#1}%
\providecommand \href@noop [0]{\@secondoftwo}%
\providecommand \href [0]{\begingroup \@sanitize@url \@href}%
\providecommand \@href[1]{\@@startlink{#1}\@@href}%
\providecommand \@@href[1]{\endgroup#1\@@endlink}%
\providecommand \@sanitize@url [0]{\catcode `\\12\catcode `\$12\catcode
  `\&12\catcode `\#12\catcode `\^12\catcode `\_12\catcode `\%12\relax}%
\providecommand \@@startlink[1]{}%
\providecommand \@@endlink[0]{}%
\providecommand \url  [0]{\begingroup\@sanitize@url \@url }%
\providecommand \@url [1]{\endgroup\@href {#1}{\urlprefix }}%
\providecommand \urlprefix  [0]{URL }%
\providecommand \Eprint [0]{\href }%
\providecommand \doibase [0]{https://doi.org/}%
\providecommand \selectlanguage [0]{\@gobble}%
\providecommand \bibinfo  [0]{\@secondoftwo}%
\providecommand \bibfield  [0]{\@secondoftwo}%
\providecommand \translation [1]{[#1]}%
\providecommand \BibitemOpen [0]{}%
\providecommand \bibitemStop [0]{}%
\providecommand \bibitemNoStop [0]{.\EOS\space}%
\providecommand \EOS [0]{\spacefactor3000\relax}%
\providecommand \BibitemShut  [1]{\csname bibitem#1\endcsname}%
\let\auto@bib@innerbib\@empty
\bibitem [{\citenamefont {Turing}(1952)}]{turing_chemical_1952}%
  \BibitemOpen
  \bibfield  {author} {\bibinfo {author} {\bibfnamefont {A.~M.}\ \bibnamefont
  {Turing}},\ }\href {https://doi.org/10.1098/rstb.1952.0012} {\bibfield
  {journal} {\bibinfo  {journal} {Philosophical Transactions of the Royal
  Society of London. Series B, Biological Sciences}\ }\textbf {\bibinfo
  {volume} {237}},\ \bibinfo {pages} {37} (\bibinfo {year} {1952})}\BibitemShut
  {NoStop}%
\bibitem [{\citenamefont {Crampin}\ \emph {et~al.}(1999)\citenamefont
  {Crampin}, \citenamefont {Gaffney},\ and\ \citenamefont
  {Maini}}]{crampin_reaction_1999}%
  \BibitemOpen
  \bibfield  {author} {\bibinfo {author} {\bibfnamefont {E.~J.}\ \bibnamefont
  {Crampin}}, \bibinfo {author} {\bibfnamefont {E.~A.}\ \bibnamefont
  {Gaffney}},\ and\ \bibinfo {author} {\bibfnamefont {P.~K.}\ \bibnamefont
  {Maini}},\ }\href {https://doi.org/10.1006/bulm.1999.0131} {\bibfield
  {journal} {\bibinfo  {journal} {Bulletin of Mathematical Biology}\ }\textbf
  {\bibinfo {volume} {61}},\ \bibinfo {pages} {1093} (\bibinfo {year}
  {1999})}\BibitemShut {NoStop}%
\bibitem [{\citenamefont {Madzvamuse}\ \emph {et~al.}(2010)\citenamefont
  {Madzvamuse}, \citenamefont {Gaffney},\ and\ \citenamefont
  {Maini}}]{madzvamuse_stability_2010}%
  \BibitemOpen
  \bibfield  {author} {\bibinfo {author} {\bibfnamefont {A.}~\bibnamefont
  {Madzvamuse}}, \bibinfo {author} {\bibfnamefont {E.~A.}\ \bibnamefont
  {Gaffney}},\ and\ \bibinfo {author} {\bibfnamefont {P.~K.}\ \bibnamefont
  {Maini}},\ }\href {https://doi.org/10.1007/s00285-009-0293-4} {\bibfield
  {journal} {\bibinfo  {journal} {Journal of Mathematical Biology}\ }\textbf
  {\bibinfo {volume} {61}},\ \bibinfo {pages} {133} (\bibinfo {year}
  {2010})}\BibitemShut {NoStop}%
\bibitem [{\citenamefont {Knobloch}\ and\ \citenamefont
  {Krechetnikov}(2015)}]{knobloch_problems_2015}%
  \BibitemOpen
  \bibfield  {author} {\bibinfo {author} {\bibfnamefont {E.}~\bibnamefont
  {Knobloch}}\ and\ \bibinfo {author} {\bibfnamefont {R.}~\bibnamefont
  {Krechetnikov}},\ }\href {https://doi.org/10.1007/s10440-014-9993-x}
  {\bibfield  {journal} {\bibinfo  {journal} {Acta Applicandae Mathematicae}\
  }\textbf {\bibinfo {volume} {137}},\ \bibinfo {pages} {123} (\bibinfo {year}
  {2015})}\BibitemShut {NoStop}%
\bibitem [{\citenamefont {Krechetnikov}(2011)}]{krechetnikov_linear_2011}%
  \BibitemOpen
  \bibfield  {author} {\bibinfo {author} {\bibfnamefont {R.}~\bibnamefont
  {Krechetnikov}},\ }\href {https://doi.org/10.4310/DPDE.2011.v8.n1.a4}
  {\bibfield  {journal} {\bibinfo  {journal} {Dynamics of Partial Differential
  Equations}\ }\textbf {\bibinfo {volume} {8}},\ \bibinfo {pages} {47}
  (\bibinfo {year} {2011})}\BibitemShut {NoStop}%
\bibitem [{\citenamefont {Peebles}(2020)}]{peebles_large-scale_2020}%
  \BibitemOpen
  \bibfield  {author} {\bibinfo {author} {\bibfnamefont {P.~J.~E.}\
  \bibnamefont {Peebles}},\ }\href
  {https://press.princeton.edu/books/paperback/9780691209838/the-large-scale-structure-of-the-universe}
  {\emph {\bibinfo {title} {The {Large}-{Scale} {Structure} of the
  {Universe}}}},\ Princeton Series in Physics\ (\bibinfo  {publisher}
  {Princeton University Press},\ \bibinfo {year} {2020})\BibitemShut {NoStop}%
\bibitem [{\citenamefont {Gierer}\ and\ \citenamefont
  {Meinhardt}(1972)}]{gierer_theory_1972}%
  \BibitemOpen
  \bibfield  {author} {\bibinfo {author} {\bibfnamefont {A.}~\bibnamefont
  {Gierer}}\ and\ \bibinfo {author} {\bibfnamefont {H.}~\bibnamefont
  {Meinhardt}},\ }\href {https://doi.org/10.1007/BF00289234} {\bibfield
  {journal} {\bibinfo  {journal} {Kybernetik}\ }\textbf {\bibinfo {volume}
  {12}},\ \bibinfo {pages} {30} (\bibinfo {year} {1972})}\BibitemShut {NoStop}%
\bibitem [{\citenamefont {Meinhardt}(1992)}]{meinhardt_pattern_1992}%
  \BibitemOpen
  \bibfield  {author} {\bibinfo {author} {\bibfnamefont {H.}~\bibnamefont
  {Meinhardt}},\ }\href {https://doi.org/10.1088/0034-4885/55/6/003} {\bibfield
   {journal} {\bibinfo  {journal} {Reports on Progress in Physics}\ }\textbf
  {\bibinfo {volume} {55}},\ \bibinfo {pages} {797} (\bibinfo {year}
  {1992})}\BibitemShut {NoStop}%
\bibitem [{\citenamefont {Crampin}\ \emph {et~al.}(2002)\citenamefont
  {Crampin}, \citenamefont {Hackborn},\ and\ \citenamefont
  {Maini}}]{crampin_pattern_2002}%
  \BibitemOpen
  \bibfield  {author} {\bibinfo {author} {\bibfnamefont {E.~J.}\ \bibnamefont
  {Crampin}}, \bibinfo {author} {\bibfnamefont {W.~W.}\ \bibnamefont
  {Hackborn}},\ and\ \bibinfo {author} {\bibfnamefont {P.~K.}\ \bibnamefont
  {Maini}},\ }\href {https://doi.org/10.1006/bulm.2002.0295} {\bibfield
  {journal} {\bibinfo  {journal} {Bulletin of Mathematical Biology}\ }\textbf
  {\bibinfo {volume} {64}},\ \bibinfo {pages} {747} (\bibinfo {year}
  {2002})}\BibitemShut {NoStop}%
\bibitem [{\citenamefont {Plaza}\ \emph {et~al.}(2004)\citenamefont {Plaza},
  \citenamefont {S\'{a}nchez-Gardu\~{n}o}, \citenamefont {Padilla},
  \citenamefont {Barrio},\ and\ \citenamefont {Maini}}]{plaza_effect_2004}%
  \BibitemOpen
  \bibfield  {author} {\bibinfo {author} {\bibfnamefont {R.}~\bibnamefont
  {Plaza}}, \bibinfo {author} {\bibfnamefont {F.}~\bibnamefont
  {S\'{a}nchez-Gardu\~{n}o}}, \bibinfo {author} {\bibfnamefont
  {P.}~\bibnamefont {Padilla}}, \bibinfo {author} {\bibfnamefont
  {R.}~\bibnamefont {Barrio}},\ and\ \bibinfo {author} {\bibfnamefont
  {P.}~\bibnamefont {Maini}},\ }\href
  {https://doi.org/10.1007/s10884-004-7834-8} {\bibfield  {journal} {\bibinfo
  {journal} {Journal of Dynamics and Differential Equations}\ }\textbf
  {\bibinfo {volume} {16}},\ \bibinfo {pages} {1093} (\bibinfo {year}
  {2004})}\BibitemShut {NoStop}%
\bibitem [{\citenamefont {Goriely}(2017)}]{goriely_mathematics_2017}%
  \BibitemOpen
  \bibfield  {author} {\bibinfo {author} {\bibfnamefont {A.}~\bibnamefont
  {Goriely}},\ }\href {https://doi.org/10.1007/978-0-387-87710-5} {\emph
  {\bibinfo {title} {The {Mathematics} and {Mechanics} of {Biological}
  {Growth}}}},\ \bibinfo {series} {Interdisciplinary {Applied} {Mathematics}},
  Vol.~\bibinfo {volume} {45}\ (\bibinfo  {publisher} {Springer New York},\
  \bibinfo {year} {2017})\BibitemShut {NoStop}%
\bibitem [{\citenamefont {Kim}\ \emph {et~al.}(2020)\citenamefont {Kim},
  \citenamefont {Yun}, \citenamefont {Yoon}, \citenamefont {Lee}, \citenamefont
  {Park},\ and\ \citenamefont {Kim}}]{kim_pattern_2020}%
  \BibitemOpen
  \bibfield  {author} {\bibinfo {author} {\bibfnamefont {H.}~\bibnamefont
  {Kim}}, \bibinfo {author} {\bibfnamefont {A.}~\bibnamefont {Yun}}, \bibinfo
  {author} {\bibfnamefont {S.}~\bibnamefont {Yoon}}, \bibinfo {author}
  {\bibfnamefont {C.}~\bibnamefont {Lee}}, \bibinfo {author} {\bibfnamefont
  {J.}~\bibnamefont {Park}},\ and\ \bibinfo {author} {\bibfnamefont
  {J.}~\bibnamefont {Kim}},\ }\href
  {https://doi.org/10.1016/j.camwa.2020.08.026} {\bibfield  {journal} {\bibinfo
   {journal} {Computers \& Mathematics with Applications}\ }\textbf {\bibinfo
  {volume} {80}},\ \bibinfo {pages} {2019} (\bibinfo {year}
  {2020})}\BibitemShut {NoStop}%
\bibitem [{\citenamefont {Ben~Tahar}\ \emph {et~al.}(2023)\citenamefont
  {Ben~Tahar}, \citenamefont {Muñoz}, \citenamefont {Shefelbine},\ and\
  \citenamefont {Comellas}}]{ben_tahar_turing_2023}%
  \BibitemOpen
  \bibfield  {author} {\bibinfo {author} {\bibfnamefont {S.}~\bibnamefont
  {Ben~Tahar}}, \bibinfo {author} {\bibfnamefont {J.~J.}\ \bibnamefont
  {Muñoz}}, \bibinfo {author} {\bibfnamefont {S.~J.}\ \bibnamefont
  {Shefelbine}},\ and\ \bibinfo {author} {\bibfnamefont {E.}~\bibnamefont
  {Comellas}},\ }\href {https://doi.org/10.1101/2023.03.29.534782} {\emph
  {\bibinfo {title} {Turing pattern prediction in three-dimensional domains:
  the role of initial conditions and growth}}},\ \bibinfo {type} {preprint}\
  (\bibinfo  {institution} {Developmental Biology},\ \bibinfo {year}
  {2023})\BibitemShut {NoStop}%
\bibitem [{\citenamefont {Duca}\ and\ \citenamefont
  {Joly}(2021)}]{duca_schrodinger_2021}%
  \BibitemOpen
  \bibfield  {author} {\bibinfo {author} {\bibfnamefont {A.}~\bibnamefont
  {Duca}}\ and\ \bibinfo {author} {\bibfnamefont {R.}~\bibnamefont {Joly}},\
  }\href {https://doi.org/10.1007/s00023-021-01020-9} {\bibfield  {journal}
  {\bibinfo  {journal} {Annales Henri Poincaré}\ }\textbf {\bibinfo {volume}
  {22}},\ \bibinfo {pages} {2029} (\bibinfo {year} {2021})}\BibitemShut
  {NoStop}%
\bibitem [{\citenamefont {Duca}\ \emph {et~al.}(2023)\citenamefont {Duca},
  \citenamefont {Joly},\ and\ \citenamefont {Turaev}}]{duca_control_2023}%
  \BibitemOpen
  \bibfield  {author} {\bibinfo {author} {\bibfnamefont {A.}~\bibnamefont
  {Duca}}, \bibinfo {author} {\bibfnamefont {R.}~\bibnamefont {Joly}},\ and\
  \bibinfo {author} {\bibfnamefont {D.}~\bibnamefont {Turaev}},\ }\href
  {https://ems.press/doi/10.4171/aihpc/86} {\bibfield  {journal} {\bibinfo
  {journal} {Annales Henri Poincaré, Analyse Non Linéaire (Online First)}\ }
  (\bibinfo {year} {2023})}\BibitemShut {NoStop}%
\bibitem [{\citenamefont {Dai}\ \emph {et~al.}(2023)\citenamefont {Dai},
  \citenamefont {Zhang}, \citenamefont {Wu},\ and\ \citenamefont
  {Tian}}]{dai_closed_2023}%
  \BibitemOpen
  \bibfield  {author} {\bibinfo {author} {\bibfnamefont {X.}~\bibnamefont
  {Dai}}, \bibinfo {author} {\bibfnamefont {H.}~\bibnamefont {Zhang}}, \bibinfo
  {author} {\bibfnamefont {Q.}~\bibnamefont {Wu}},\ and\ \bibinfo {author}
  {\bibfnamefont {S.}~\bibnamefont {Tian}},\ }\href
  {https://doi.org/10.1002/oca.2779} {\bibfield  {journal} {\bibinfo  {journal}
  {Optimal Control Applications and Methods}\ }\textbf {\bibinfo {volume}
  {44}},\ \bibinfo {pages} {1200} (\bibinfo {year} {2023})}\BibitemShut
  {NoStop}%
\bibitem [{\citenamefont {Benoît}(1991)}]{benoit_dynamic_1991}%
  \BibitemOpen
  \bibinfo {editor} {\bibfnamefont {E.}~\bibnamefont {Benoît}},\ ed.,\ \href
  {https://link.springer.com/book/10.1007/BFb0085019} {\emph {\bibinfo {title}
  {Dynamic Bifurcations}}},\ \bibinfo {series} {Lecture Notes in Mathematics}\
  No.\ \bibinfo {number} {1493}\ (\bibinfo  {publisher} {Springer-Verlag},\
  \bibinfo {address} {Berlin},\ \bibinfo {year} {1991})\BibitemShut {NoStop}%
\bibitem [{\citenamefont {Ashwin}\ \emph {et~al.}(2012)\citenamefont {Ashwin},
  \citenamefont {Wieczorek}, \citenamefont {Vitolo},\ and\ \citenamefont
  {Cox}}]{ashwin_tipping_2012}%
  \BibitemOpen
  \bibfield  {author} {\bibinfo {author} {\bibfnamefont {P.}~\bibnamefont
  {Ashwin}}, \bibinfo {author} {\bibfnamefont {S.}~\bibnamefont {Wieczorek}},
  \bibinfo {author} {\bibfnamefont {R.}~\bibnamefont {Vitolo}},\ and\ \bibinfo
  {author} {\bibfnamefont {P.}~\bibnamefont {Cox}},\ }\href
  {https://doi.org/10.1098/rsta.2011.0306} {\bibfield  {journal} {\bibinfo
  {journal} {Philosophical Transactions of the Royal Society A: Mathematical,
  Physical and Engineering Sciences}\ }\textbf {\bibinfo {volume} {370}},\
  \bibinfo {pages} {1166} (\bibinfo {year} {2012})}\BibitemShut {NoStop}%
\bibitem [{\citenamefont {Premraj}\ \emph {et~al.}(2016)\citenamefont
  {Premraj}, \citenamefont {Suresh}, \citenamefont {Banerjee},\ and\
  \citenamefont {Thamilmaran}}]{premraj_experimental_2016}%
  \BibitemOpen
  \bibfield  {author} {\bibinfo {author} {\bibfnamefont {D.}~\bibnamefont
  {Premraj}}, \bibinfo {author} {\bibfnamefont {K.}~\bibnamefont {Suresh}},
  \bibinfo {author} {\bibfnamefont {T.}~\bibnamefont {Banerjee}},\ and\
  \bibinfo {author} {\bibfnamefont {K.}~\bibnamefont {Thamilmaran}},\ }\href
  {https://doi.org/10.1016/j.cnsns.2016.01.012} {\bibfield  {journal} {\bibinfo
   {journal} {Communications in Nonlinear Science and Numerical Simulation}\
  }\textbf {\bibinfo {volume} {37}},\ \bibinfo {pages} {212} (\bibinfo {year}
  {2016})}\BibitemShut {NoStop}%
\bibitem [{\citenamefont {Premraj}\ \emph {et~al.}(2019)\citenamefont
  {Premraj}, \citenamefont {Suresh},\ and\ \citenamefont
  {Thamilmaran}}]{premraj_effect_2019}%
  \BibitemOpen
  \bibfield  {author} {\bibinfo {author} {\bibfnamefont {D.}~\bibnamefont
  {Premraj}}, \bibinfo {author} {\bibfnamefont {K.}~\bibnamefont {Suresh}},\
  and\ \bibinfo {author} {\bibfnamefont {K.}~\bibnamefont {Thamilmaran}},\
  }\href {https://doi.org/10.1063/1.5123417} {\bibfield  {journal} {\bibinfo
  {journal} {Chaos: An Interdisciplinary Journal of Nonlinear Science}\
  }\textbf {\bibinfo {volume} {29}},\ \bibinfo {pages} {123127} (\bibinfo
  {year} {2019})}\BibitemShut {NoStop}%
\bibitem [{\citenamefont {Varshney}\ \emph {et~al.}(2020)\citenamefont
  {Varshney}, \citenamefont {Kumarasamy}, \citenamefont {Biswal},\ and\
  \citenamefont {Prasad}}]{varshney_bifurcation_2020}%
  \BibitemOpen
  \bibfield  {author} {\bibinfo {author} {\bibfnamefont {V.}~\bibnamefont
  {Varshney}}, \bibinfo {author} {\bibfnamefont {S.}~\bibnamefont
  {Kumarasamy}}, \bibinfo {author} {\bibfnamefont {B.}~\bibnamefont {Biswal}},\
  and\ \bibinfo {author} {\bibfnamefont {A.}~\bibnamefont {Prasad}},\ }\href
  {https://doi.org/10.1140/epjst/e2020-900192-x} {\bibfield  {journal}
  {\bibinfo  {journal} {The European Physical Journal Special Topics}\ }\textbf
  {\bibinfo {volume} {229}},\ \bibinfo {pages} {2307} (\bibinfo {year}
  {2020})}\BibitemShut {NoStop}%
\bibitem [{\citenamefont {Talla~Mbé}\ and\ \citenamefont
  {Woafo}(2020)}]{talla_mbe_study_2020}%
  \BibitemOpen
  \bibfield  {author} {\bibinfo {author} {\bibfnamefont {J.~H.}\ \bibnamefont
  {Talla~Mbé}}\ and\ \bibinfo {author} {\bibfnamefont {P.}~\bibnamefont
  {Woafo}},\ }\href {https://doi.org/10.1063/5.0004638} {\bibfield  {journal}
  {\bibinfo  {journal} {Chaos: An Interdisciplinary Journal of Nonlinear
  Science}\ }\textbf {\bibinfo {volume} {30}},\ \bibinfo {pages} {093130}
  (\bibinfo {year} {2020})}\BibitemShut {NoStop}%
\bibitem [{\citenamefont {Rinzel}\ and\ \citenamefont
  {Baer}(1988)}]{rinzel_threshold_1988}%
  \BibitemOpen
  \bibfield  {author} {\bibinfo {author} {\bibfnamefont {J.}~\bibnamefont
  {Rinzel}}\ and\ \bibinfo {author} {\bibfnamefont {S.~M.}\ \bibnamefont
  {Baer}},\ }\href {https://doi.org/10.1016/S0006-3495(88)82988-6} {\bibfield
  {journal} {\bibinfo  {journal} {Biophysical Journal}\ }\textbf {\bibinfo
  {volume} {54}},\ \bibinfo {pages} {551} (\bibinfo {year} {1988})}\BibitemShut
  {NoStop}%
\bibitem [{\citenamefont {Ahlers}\ \emph {et~al.}(1981)\citenamefont {Ahlers},
  \citenamefont {Cross}, \citenamefont {Hohenberg},\ and\ \citenamefont
  {Safran}}]{ahlers_amplitude_1981}%
  \BibitemOpen
  \bibfield  {author} {\bibinfo {author} {\bibfnamefont {G.}~\bibnamefont
  {Ahlers}}, \bibinfo {author} {\bibfnamefont {M.~C.}\ \bibnamefont {Cross}},
  \bibinfo {author} {\bibfnamefont {P.~C.}\ \bibnamefont {Hohenberg}},\ and\
  \bibinfo {author} {\bibfnamefont {S.}~\bibnamefont {Safran}},\ }\href
  {https://doi.org/10.1017/S0022112081000761} {\bibfield  {journal} {\bibinfo
  {journal} {Journal of Fluid Mechanics}\ }\textbf {\bibinfo {volume} {110}},\
  \bibinfo {pages} {297} (\bibinfo {year} {1981})}\BibitemShut {NoStop}%
\bibitem [{\citenamefont {Laplante}\ \emph {et~al.}(1991)\citenamefont
  {Laplante}, \citenamefont {Erneux},\ and\ \citenamefont
  {Georgiou}}]{laplante_jump_1991}%
  \BibitemOpen
  \bibfield  {author} {\bibinfo {author} {\bibfnamefont {J.~P.}\ \bibnamefont
  {Laplante}}, \bibinfo {author} {\bibfnamefont {T.}~\bibnamefont {Erneux}},\
  and\ \bibinfo {author} {\bibfnamefont {M.}~\bibnamefont {Georgiou}},\ }\href
  {https://doi.org/10.1063/1.460352} {\bibfield  {journal} {\bibinfo  {journal}
  {The Journal of Chemical Physics}\ }\textbf {\bibinfo {volume} {94}},\
  \bibinfo {pages} {371} (\bibinfo {year} {1991})}\BibitemShut {NoStop}%
\bibitem [{\citenamefont {Strizhak}\ and\ \citenamefont
  {Menzinger}(1996)}]{strizhak_slow_1996}%
  \BibitemOpen
  \bibfield  {author} {\bibinfo {author} {\bibfnamefont {P.}~\bibnamefont
  {Strizhak}}\ and\ \bibinfo {author} {\bibfnamefont {M.}~\bibnamefont
  {Menzinger}},\ }\href {https://doi.org/10.1063/1.472860} {\bibfield
  {journal} {\bibinfo  {journal} {The Journal of Chemical Physics}\ }\textbf
  {\bibinfo {volume} {105}},\ \bibinfo {pages} {10905} (\bibinfo {year}
  {1996})}\BibitemShut {NoStop}%
\bibitem [{\citenamefont {De~Maesschalck}\ \emph {et~al.}(2009)\citenamefont
  {De~Maesschalck}, \citenamefont {Kaper},\ and\ \citenamefont
  {Popović}}]{de_maesschalck_canards_2009}%
  \BibitemOpen
  \bibfield  {author} {\bibinfo {author} {\bibfnamefont {P.}~\bibnamefont
  {De~Maesschalck}}, \bibinfo {author} {\bibfnamefont {T.~J.}\ \bibnamefont
  {Kaper}},\ and\ \bibinfo {author} {\bibfnamefont {N.}~\bibnamefont
  {Popović}},\ }\href {https://doi.org/10.57262/ade/1355863335} {\bibfield
  {journal} {\bibinfo  {journal} {Advances in Differential Equations}\ }\textbf
  {\bibinfo {volume} {14}},\ \bibinfo {pages} {943} (\bibinfo {year}
  {2009})}\BibitemShut {NoStop}%
\bibitem [{\citenamefont {Vasil}\ and\ \citenamefont
  {Proctor}(2011)}]{vasil_dynamic_2011}%
  \BibitemOpen
  \bibfield  {author} {\bibinfo {author} {\bibfnamefont {G.~M.}\ \bibnamefont
  {Vasil}}\ and\ \bibinfo {author} {\bibfnamefont {M.~R.~E.}\ \bibnamefont
  {Proctor}},\ }\href {https://doi.org/10.1017/jfm.2011.284} {\bibfield
  {journal} {\bibinfo  {journal} {Journal of Fluid Mechanics}\ }\textbf
  {\bibinfo {volume} {686}},\ \bibinfo {pages} {77} (\bibinfo {year}
  {2011})}\BibitemShut {NoStop}%
\bibitem [{\citenamefont {Premraj}\ \emph {et~al.}(2017)\citenamefont
  {Premraj}, \citenamefont {Suresh}, \citenamefont {Banerjee},\ and\
  \citenamefont {Thamilmaran}}]{premraj_control_2017}%
  \BibitemOpen
  \bibfield  {author} {\bibinfo {author} {\bibfnamefont {D.}~\bibnamefont
  {Premraj}}, \bibinfo {author} {\bibfnamefont {K.}~\bibnamefont {Suresh}},
  \bibinfo {author} {\bibfnamefont {T.}~\bibnamefont {Banerjee}},\ and\
  \bibinfo {author} {\bibfnamefont {K.}~\bibnamefont {Thamilmaran}},\ }\href
  {https://doi.org/10.1063/1.4973237} {\bibfield  {journal} {\bibinfo
  {journal} {Chaos: An Interdisciplinary Journal of Nonlinear Science}\
  }\textbf {\bibinfo {volume} {27}},\ \bibinfo {pages} {013104} (\bibinfo
  {year} {2017})}\BibitemShut {NoStop}%
\bibitem [{\citenamefont {Yu}\ and\ \citenamefont
  {Christov}(2023)}]{yu_delayed_2023}%
  \BibitemOpen
  \bibfield  {author} {\bibinfo {author} {\bibfnamefont {Z.}~\bibnamefont
  {Yu}}\ and\ \bibinfo {author} {\bibfnamefont {I.~C.}\ \bibnamefont
  {Christov}},\ }\href {https://doi.org/10.1103/PhysRevE.107.055102} {\bibfield
   {journal} {\bibinfo  {journal} {Physical Review E}\ }\textbf {\bibinfo
  {volume} {107}},\ \bibinfo {pages} {055102} (\bibinfo {year}
  {2023})}\BibitemShut {NoStop}%
\bibitem [{\citenamefont {Mandel}\ and\ \citenamefont
  {Erneux}(1987)}]{mandel_slow_1987}%
  \BibitemOpen
  \bibfield  {author} {\bibinfo {author} {\bibfnamefont {P.}~\bibnamefont
  {Mandel}}\ and\ \bibinfo {author} {\bibfnamefont {T.}~\bibnamefont
  {Erneux}},\ }\href {https://doi.org/10.1007/BF01009533} {\bibfield  {journal}
  {\bibinfo  {journal} {Journal of Statistical Physics}\ }\textbf {\bibinfo
  {volume} {48}},\ \bibinfo {pages} {1059} (\bibinfo {year}
  {1987})}\BibitemShut {NoStop}%
\bibitem [{\citenamefont {Baer}\ \emph {et~al.}(1989)\citenamefont {Baer},
  \citenamefont {Erneux},\ and\ \citenamefont {Rinzel}}]{baer_slow_1989}%
  \BibitemOpen
  \bibfield  {author} {\bibinfo {author} {\bibfnamefont {S.~M.}\ \bibnamefont
  {Baer}}, \bibinfo {author} {\bibfnamefont {T.}~\bibnamefont {Erneux}},\ and\
  \bibinfo {author} {\bibfnamefont {J.}~\bibnamefont {Rinzel}},\ }\href
  {https://doi.org/10.1137/0149003} {\bibfield  {journal} {\bibinfo  {journal}
  {SIAM Journal on Applied Mathematics}\ }\textbf {\bibinfo {volume} {49}},\
  \bibinfo {pages} {55} (\bibinfo {year} {1989})}\BibitemShut {NoStop}%
\bibitem [{\citenamefont {Lythe}(1996)}]{lythe_domain_1996}%
  \BibitemOpen
  \bibfield  {author} {\bibinfo {author} {\bibfnamefont {G.~D.}\ \bibnamefont
  {Lythe}},\ }\href {https://doi.org/10.1103/PhysRevE.53.R4271} {\bibfield
  {journal} {\bibinfo  {journal} {Physical Review E}\ }\textbf {\bibinfo
  {volume} {53}},\ \bibinfo {pages} {R4271} (\bibinfo {year}
  {1996})}\BibitemShut {NoStop}%
\bibitem [{\citenamefont {Haberman}(2001)}]{haberman_slow_2001}%
  \BibitemOpen
  \bibfield  {author} {\bibinfo {author} {\bibfnamefont {R.}~\bibnamefont
  {Haberman}},\ }\href {https://doi.org/10.1137/S0036139900373836} {\bibfield
  {journal} {\bibinfo  {journal} {SIAM Journal on Applied Mathematics}\
  }\textbf {\bibinfo {volume} {62}},\ \bibinfo {pages} {488} (\bibinfo {year}
  {2001})}\BibitemShut {NoStop}%
\bibitem [{\citenamefont {Diminnie}\ and\ \citenamefont
  {Haberman}(2002)}]{diminnie_slow_2002}%
  \BibitemOpen
  \bibfield  {author} {\bibinfo {author} {\bibfnamefont {D.~C.}\ \bibnamefont
  {Diminnie}}\ and\ \bibinfo {author} {\bibfnamefont {R.}~\bibnamefont
  {Haberman}},\ }\href {https://doi.org/10.1016/S0167-2789(01)00373-6}
  {\bibfield  {journal} {\bibinfo  {journal} {Physica D: Nonlinear Phenomena}\
  }\textbf {\bibinfo {volume} {162}},\ \bibinfo {pages} {34} (\bibinfo {year}
  {2002})}\BibitemShut {NoStop}%
\bibitem [{\citenamefont {Ng}\ \emph {et~al.}(2003)\citenamefont {Ng},
  \citenamefont {Rand},\ and\ \citenamefont {O'Neil}}]{ng_slow_2003}%
  \BibitemOpen
  \bibfield  {author} {\bibinfo {author} {\bibfnamefont {L.}~\bibnamefont
  {Ng}}, \bibinfo {author} {\bibfnamefont {R.}~\bibnamefont {Rand}},\ and\
  \bibinfo {author} {\bibfnamefont {M.}~\bibnamefont {O'Neil}},\ }\href
  {https://doi.org/10.1177/1077546303009006004} {\bibfield  {journal} {\bibinfo
   {journal} {Journal of Vibration and Control}\ }\textbf {\bibinfo {volume}
  {9}},\ \bibinfo {pages} {685} (\bibinfo {year} {2003})}\BibitemShut {NoStop}%
\bibitem [{\citenamefont {Neishtadt}(2009)}]{neishtadt_stability_2009}%
  \BibitemOpen
  \bibfield  {author} {\bibinfo {author} {\bibfnamefont {A.}~\bibnamefont
  {Neishtadt}},\ }\href {https://doi.org/10.3934/dcdss.2009.2.897} {\bibfield
  {journal} {\bibinfo  {journal} {Discrete \& Continuous Dynamical Systems -
  S}\ }\textbf {\bibinfo {volume} {2}},\ \bibinfo {pages} {897} (\bibinfo
  {year} {2009})}\BibitemShut {NoStop}%
\bibitem [{\citenamefont {Park}\ \emph {et~al.}(2011)\citenamefont {Park},
  \citenamefont {Do},\ and\ \citenamefont {Lopez}}]{park_slow_2011}%
  \BibitemOpen
  \bibfield  {author} {\bibinfo {author} {\bibfnamefont {Y.}~\bibnamefont
  {Park}}, \bibinfo {author} {\bibfnamefont {Y.}~\bibnamefont {Do}},\ and\
  \bibinfo {author} {\bibfnamefont {J.~M.}\ \bibnamefont {Lopez}},\ }\href
  {https://doi.org/10.1103/PhysRevE.84.056604} {\bibfield  {journal} {\bibinfo
  {journal} {Physical Review E}\ }\textbf {\bibinfo {volume} {84}},\ \bibinfo
  {pages} {056604} (\bibinfo {year} {2011})}\BibitemShut {NoStop}%
\bibitem [{\citenamefont {Han}\ \emph {et~al.}(2014)\citenamefont {Han},
  \citenamefont {Bi}, \citenamefont {Zhang},\ and\ \citenamefont
  {Yu}}]{han_delayed_2014}%
  \BibitemOpen
  \bibfield  {author} {\bibinfo {author} {\bibfnamefont {X.}~\bibnamefont
  {Han}}, \bibinfo {author} {\bibfnamefont {Q.}~\bibnamefont {Bi}}, \bibinfo
  {author} {\bibfnamefont {C.}~\bibnamefont {Zhang}},\ and\ \bibinfo {author}
  {\bibfnamefont {Y.}~\bibnamefont {Yu}},\ }\href
  {https://doi.org/10.1142/S0218127414500989} {\bibfield  {journal} {\bibinfo
  {journal} {International Journal of Bifurcation and Chaos}\ }\textbf
  {\bibinfo {volume} {24}},\ \bibinfo {pages} {1450098} (\bibinfo {year}
  {2014})}\BibitemShut {NoStop}%
\bibitem [{\citenamefont {Knobloch}\ and\ \citenamefont
  {Krechetnikov}(2014)}]{knobloch_stability_2014}%
  \BibitemOpen
  \bibfield  {author} {\bibinfo {author} {\bibfnamefont {E.}~\bibnamefont
  {Knobloch}}\ and\ \bibinfo {author} {\bibfnamefont {R.}~\bibnamefont
  {Krechetnikov}},\ }\href {https://doi.org/10.1007/s00332-014-9197-6}
  {\bibfield  {journal} {\bibinfo  {journal} {Journal of Nonlinear Science}\
  }\textbf {\bibinfo {volume} {24}},\ \bibinfo {pages} {493} (\bibinfo {year}
  {2014})}\BibitemShut {NoStop}%
\bibitem [{\citenamefont {Krechetnikov}\ and\ \citenamefont
  {Knobloch}(2017)}]{krechetnikov_stability_2017}%
  \BibitemOpen
  \bibfield  {author} {\bibinfo {author} {\bibfnamefont {R.}~\bibnamefont
  {Krechetnikov}}\ and\ \bibinfo {author} {\bibfnamefont {E.}~\bibnamefont
  {Knobloch}},\ }\href {https://doi.org/10.1016/j.physd.2016.10.003} {\bibfield
   {journal} {\bibinfo  {journal} {Physica D: Nonlinear Phenomena}\ }\textbf
  {\bibinfo {volume} {342}},\ \bibinfo {pages} {16} (\bibinfo {year}
  {2017})}\BibitemShut {NoStop}%
\bibitem [{\citenamefont {van Saarloos}(2003)}]{van_saarloos_front_2003}%
  \BibitemOpen
  \bibfield  {author} {\bibinfo {author} {\bibfnamefont {W.}~\bibnamefont {van
  Saarloos}},\ }\href {https://doi.org/10.1016/j.physrep.2003.08.001}
  {\bibfield  {journal} {\bibinfo  {journal} {Physics Reports}\ }\textbf
  {\bibinfo {volume} {386}},\ \bibinfo {pages} {29} (\bibinfo {year}
  {2003})}\BibitemShut {NoStop}%
\bibitem [{\citenamefont {Ahlers}\ and\ \citenamefont
  {Cannell}(1983)}]{ahlers_vortex-front_1983}%
  \BibitemOpen
  \bibfield  {author} {\bibinfo {author} {\bibfnamefont {G.}~\bibnamefont
  {Ahlers}}\ and\ \bibinfo {author} {\bibfnamefont {D.~S.}\ \bibnamefont
  {Cannell}},\ }\href {https://doi.org/10.1103/PhysRevLett.50.1583} {\bibfield
  {journal} {\bibinfo  {journal} {Physical Review Letters}\ }\textbf {\bibinfo
  {volume} {50}},\ \bibinfo {pages} {1583} (\bibinfo {year}
  {1983})}\BibitemShut {NoStop}%
\bibitem [{\citenamefont {Powers}\ \emph {et~al.}(1998)\citenamefont {Powers},
  \citenamefont {Zhang}, \citenamefont {Goldstein},\ and\ \citenamefont
  {Stone}}]{powers_propagation_1998}%
  \BibitemOpen
  \bibfield  {author} {\bibinfo {author} {\bibfnamefont {T.~R.}\ \bibnamefont
  {Powers}}, \bibinfo {author} {\bibfnamefont {D.}~\bibnamefont {Zhang}},
  \bibinfo {author} {\bibfnamefont {R.~E.}\ \bibnamefont {Goldstein}},\ and\
  \bibinfo {author} {\bibfnamefont {H.~A.}\ \bibnamefont {Stone}},\ }\href
  {https://doi.org/10.1063/1.869650} {\bibfield  {journal} {\bibinfo  {journal}
  {Physics of Fluids}\ }\textbf {\bibinfo {volume} {10}},\ \bibinfo {pages}
  {1052} (\bibinfo {year} {1998})}\BibitemShut {NoStop}%
\bibitem [{\citenamefont {Maini}\ \emph {et~al.}(2004)\citenamefont {Maini},
  \citenamefont {McElwain},\ and\ \citenamefont
  {Leavesley}}]{maini_travelling_2004}%
  \BibitemOpen
  \bibfield  {author} {\bibinfo {author} {\bibfnamefont {P.}~\bibnamefont
  {Maini}}, \bibinfo {author} {\bibfnamefont {D.}~\bibnamefont {McElwain}},\
  and\ \bibinfo {author} {\bibfnamefont {D.}~\bibnamefont {Leavesley}},\ }\href
  {https://doi.org/10.1016/S0893-9659(04)90128-0} {\bibfield  {journal}
  {\bibinfo  {journal} {Applied Mathematics Letters}\ }\textbf {\bibinfo
  {volume} {17}},\ \bibinfo {pages} {575} (\bibinfo {year} {2004})}\BibitemShut
  {NoStop}%
\bibitem [{\citenamefont {Stoop}\ and\ \citenamefont
  {Dunkel}(2018)}]{stoop_defect_2018}%
  \BibitemOpen
  \bibfield  {author} {\bibinfo {author} {\bibfnamefont {N.}~\bibnamefont
  {Stoop}}\ and\ \bibinfo {author} {\bibfnamefont {J.}~\bibnamefont {Dunkel}},\
  }\href {https://doi.org/10.1039/C7SM02233F} {\bibfield  {journal} {\bibinfo
  {journal} {Soft Matter}\ }\textbf {\bibinfo {volume} {14}},\ \bibinfo {pages}
  {2329} (\bibinfo {year} {2018})}\BibitemShut {NoStop}%
\bibitem [{\citenamefont {Rietkerk}\ \emph {et~al.}(2021)\citenamefont
  {Rietkerk}, \citenamefont {Bastiaansen}, \citenamefont {Banerjee},
  \citenamefont {van~de Koppel}, \citenamefont {Baudena},\ and\ \citenamefont
  {Doelman}}]{rietkerk_evasion_2021}%
  \BibitemOpen
  \bibfield  {author} {\bibinfo {author} {\bibfnamefont {M.}~\bibnamefont
  {Rietkerk}}, \bibinfo {author} {\bibfnamefont {R.}~\bibnamefont
  {Bastiaansen}}, \bibinfo {author} {\bibfnamefont {S.}~\bibnamefont
  {Banerjee}}, \bibinfo {author} {\bibfnamefont {J.}~\bibnamefont {van~de
  Koppel}}, \bibinfo {author} {\bibfnamefont {M.}~\bibnamefont {Baudena}},\
  and\ \bibinfo {author} {\bibfnamefont {A.}~\bibnamefont {Doelman}},\ }\href
  {https://doi.org/10.1126/science.abj0359} {\bibfield  {journal} {\bibinfo
  {journal} {Science}\ }\textbf {\bibinfo {volume} {374}},\ \bibinfo {pages}
  {eabj0359} (\bibinfo {year} {2021})}\BibitemShut {NoStop}%
\bibitem [{\citenamefont {Goh}\ \emph {et~al.}(2023)\citenamefont {Goh},
  \citenamefont {Kaper}, \citenamefont {Scheel},\ and\ \citenamefont
  {Vo}}]{goh_fronts_2022}%
  \BibitemOpen
  \bibfield  {author} {\bibinfo {author} {\bibfnamefont {R.}~\bibnamefont
  {Goh}}, \bibinfo {author} {\bibfnamefont {T.~J.}\ \bibnamefont {Kaper}},
  \bibinfo {author} {\bibfnamefont {A.}~\bibnamefont {Scheel}},\ and\ \bibinfo
  {author} {\bibfnamefont {T.}~\bibnamefont {Vo}},\ }\href
  {https://doi.org/10.1137/22M1541812} {\bibfield  {journal} {\bibinfo
  {journal} {SIAM Journal on Applied Dynamical Systems}\ }\textbf {\bibinfo
  {volume} {22}},\ \bibinfo {pages} {2312} (\bibinfo {year}
  {2023})}\BibitemShut {NoStop}%
\bibitem [{\citenamefont {Goh}\ and\ \citenamefont
  {Scheel}(2023)}]{goh_growing_2023}%
  \BibitemOpen
  \bibfield  {author} {\bibinfo {author} {\bibfnamefont {R.}~\bibnamefont
  {Goh}}\ and\ \bibinfo {author} {\bibfnamefont {A.}~\bibnamefont {Scheel}},\
  }\href {https://doi.org/10.1088/1361-6544/acf265} {\bibfield  {journal}
  {\bibinfo  {journal} {Nonlinearity}\ }\textbf {\bibinfo {volume} {36}},\
  \bibinfo {pages} {R1} (\bibinfo {year} {2023})}\BibitemShut {NoStop}%
\bibitem [{\citenamefont {Segel}(1969)}]{segel_distant_1969}%
  \BibitemOpen
  \bibfield  {author} {\bibinfo {author} {\bibfnamefont {L.~A.}\ \bibnamefont
  {Segel}},\ }\href {https://doi.org/10.1017/S0022112069000127} {\bibfield
  {journal} {\bibinfo  {journal} {Journal of Fluid Mechanics}\ }\textbf
  {\bibinfo {volume} {38}},\ \bibinfo {pages} {203} (\bibinfo {year}
  {1969})}\BibitemShut {NoStop}%
\bibitem [{\citenamefont {Newell}\ and\ \citenamefont
  {Whitehead}(1969)}]{newell_finite_1969}%
  \BibitemOpen
  \bibfield  {author} {\bibinfo {author} {\bibfnamefont {A.~C.}\ \bibnamefont
  {Newell}}\ and\ \bibinfo {author} {\bibfnamefont {J.~A.}\ \bibnamefont
  {Whitehead}},\ }\href {https://doi.org/10.1017/S0022112069000176} {\bibfield
  {journal} {\bibinfo  {journal} {Journal of Fluid Mechanics}\ }\textbf
  {\bibinfo {volume} {38}},\ \bibinfo {pages} {279} (\bibinfo {year}
  {1969})}\BibitemShut {NoStop}%
\bibitem [{\citenamefont {Ahlers}\ \emph {et~al.}(1986)\citenamefont {Ahlers},
  \citenamefont {Cannell}, \citenamefont {Dominguez-Lerma},\ and\ \citenamefont
  {Heinrichs}}]{ahlers_wavenumber_1986}%
  \BibitemOpen
  \bibfield  {author} {\bibinfo {author} {\bibfnamefont {G.}~\bibnamefont
  {Ahlers}}, \bibinfo {author} {\bibfnamefont {D.~S.}\ \bibnamefont {Cannell}},
  \bibinfo {author} {\bibfnamefont {M.~A.}\ \bibnamefont {Dominguez-Lerma}},\
  and\ \bibinfo {author} {\bibfnamefont {R.}~\bibnamefont {Heinrichs}},\ }\href
  {https://doi.org/10.1016/0167-2789(86)90129-6} {\bibfield  {journal}
  {\bibinfo  {journal} {Physica D: Nonlinear Phenomena}\ }\textbf {\bibinfo
  {volume} {23}},\ \bibinfo {pages} {202} (\bibinfo {year} {1986})}\BibitemShut
  {NoStop}%
\bibitem [{\citenamefont {Cross}\ and\ \citenamefont
  {Hohenberg}(1993)}]{cross_pattern_1993}%
  \BibitemOpen
  \bibfield  {author} {\bibinfo {author} {\bibfnamefont {M.~C.}\ \bibnamefont
  {Cross}}\ and\ \bibinfo {author} {\bibfnamefont {P.~C.}\ \bibnamefont
  {Hohenberg}},\ }\href {https://doi.org/10.1103/RevModPhys.65.851} {\bibfield
  {journal} {\bibinfo  {journal} {Reviews of Modern Physics}\ }\textbf
  {\bibinfo {volume} {65}},\ \bibinfo {pages} {851} (\bibinfo {year}
  {1993})}\BibitemShut {NoStop}%
\bibitem [{\citenamefont {Kramer}\ and\ \citenamefont
  {Hohenberg}(1984)}]{kramer_effects_1984}%
  \BibitemOpen
  \bibfield  {author} {\bibinfo {author} {\bibfnamefont {L.}~\bibnamefont
  {Kramer}}\ and\ \bibinfo {author} {\bibfnamefont {P.}~\bibnamefont
  {Hohenberg}},\ }\href {https://doi.org/10.1016/0167-2789(84)90136-2}
  {\bibfield  {journal} {\bibinfo  {journal} {Physica D: Nonlinear Phenomena}\
  }\textbf {\bibinfo {volume} {13}},\ \bibinfo {pages} {357} (\bibinfo {year}
  {1984})}\BibitemShut {NoStop}%
\bibitem [{\citenamefont {Tuckerman}\ and\ \citenamefont
  {Barkley}(1990)}]{tuckerman_bifurcation_1990}%
  \BibitemOpen
  \bibfield  {author} {\bibinfo {author} {\bibfnamefont {L.~S.}\ \bibnamefont
  {Tuckerman}}\ and\ \bibinfo {author} {\bibfnamefont {D.}~\bibnamefont
  {Barkley}},\ }\href {https://doi.org/10.1016/0167-2789(90)90113-4} {\bibfield
   {journal} {\bibinfo  {journal} {Physica D: Nonlinear Phenomena}\ }\textbf
  {\bibinfo {volume} {46}},\ \bibinfo {pages} {57} (\bibinfo {year}
  {1990})}\BibitemShut {NoStop}%
\bibitem [{\citenamefont {Eckmann}\ and\ \citenamefont
  {Gallay}(1993)}]{eckmann_front_1993}%
  \BibitemOpen
  \bibfield  {author} {\bibinfo {author} {\bibfnamefont {J.~P.}\ \bibnamefont
  {Eckmann}}\ and\ \bibinfo {author} {\bibfnamefont {T.}~\bibnamefont
  {Gallay}},\ }\href {https://doi.org/10.1007/BF02098298} {\bibfield  {journal}
  {\bibinfo  {journal} {Communications in Mathematical Physics}\ }\textbf
  {\bibinfo {volume} {152}},\ \bibinfo {pages} {221} (\bibinfo {year}
  {1993})}\BibitemShut {NoStop}%
\bibitem [{\citenamefont {Bridges}\ and\ \citenamefont
  {Rowlands}(1994)}]{bridges_instability_1994}%
  \BibitemOpen
  \bibfield  {author} {\bibinfo {author} {\bibfnamefont {T.~J.}\ \bibnamefont
  {Bridges}}\ and\ \bibinfo {author} {\bibfnamefont {G.}~\bibnamefont
  {Rowlands}},\ }\href {https://www.jstor.org/stable/52493} {\bibfield
  {journal} {\bibinfo  {journal} {Proceedings: Mathematical and Physical
  Sciences}\ }\textbf {\bibinfo {volume} {444}},\ \bibinfo {pages} {347}
  (\bibinfo {year} {1994})}\BibitemShut {NoStop}%
\bibitem [{\citenamefont {Doelman}\ \emph {et~al.}(1995)\citenamefont
  {Doelman}, \citenamefont {Gardner},\ and\ \citenamefont
  {Jones}}]{doelman_instability_1995}%
  \BibitemOpen
  \bibfield  {author} {\bibinfo {author} {\bibfnamefont {A.}~\bibnamefont
  {Doelman}}, \bibinfo {author} {\bibfnamefont {R.~A.}\ \bibnamefont
  {Gardner}},\ and\ \bibinfo {author} {\bibfnamefont {C.~K. R.~T.}\
  \bibnamefont {Jones}},\ }\href {https://doi.org/10.1017/S0308210500032649}
  {\bibfield  {journal} {\bibinfo  {journal} {Proceedings of the Royal Society
  of Edinburgh Section A: Mathematics}\ }\textbf {\bibinfo {volume} {125}},\
  \bibinfo {pages} {501} (\bibinfo {year} {1995})}\BibitemShut {NoStop}%
\bibitem [{\citenamefont {Eckmann}\ \emph {et~al.}(1995)\citenamefont
  {Eckmann}, \citenamefont {Gallay},\ and\ \citenamefont
  {Wayne}}]{eckmann_phase_1995}%
  \BibitemOpen
  \bibfield  {author} {\bibinfo {author} {\bibfnamefont {J.~P.}\ \bibnamefont
  {Eckmann}}, \bibinfo {author} {\bibfnamefont {T.}~\bibnamefont {Gallay}},\
  and\ \bibinfo {author} {\bibfnamefont {C.~E.}\ \bibnamefont {Wayne}},\ }\href
  {https://doi.org/10.1088/0951-7715/8/6/004} {\bibfield  {journal} {\bibinfo
  {journal} {Nonlinearity}\ }\textbf {\bibinfo {volume} {8}},\ \bibinfo {pages}
  {943} (\bibinfo {year} {1995})}\BibitemShut {NoStop}%
\bibitem [{\citenamefont {Cross}\ and\ \citenamefont
  {Greenside}(2009)}]{cross_pattern_2009}%
  \BibitemOpen
  \bibfield  {author} {\bibinfo {author} {\bibfnamefont {M.}~\bibnamefont
  {Cross}}\ and\ \bibinfo {author} {\bibfnamefont {H.}~\bibnamefont
  {Greenside}},\ }\href {https://doi.org/10.1017/CBO9780511627200} {\emph
  {\bibinfo {title} {Pattern {Formation} and {Dynamics} in {Nonequilibrium}
  {Systems}}}},\ \bibinfo {edition} {1st}\ ed.\ (\bibinfo  {publisher}
  {Cambridge University Press},\ \bibinfo {year} {2009})\BibitemShut {NoStop}%
\bibitem [{\citenamefont {Goh}\ \emph {et~al.}(2016)\citenamefont {Goh},
  \citenamefont {Beekie}, \citenamefont {Matthias}, \citenamefont {Nunley},\
  and\ \citenamefont {Scheel}}]{goh_universal_2016}%
  \BibitemOpen
  \bibfield  {author} {\bibinfo {author} {\bibfnamefont {R.}~\bibnamefont
  {Goh}}, \bibinfo {author} {\bibfnamefont {R.}~\bibnamefont {Beekie}},
  \bibinfo {author} {\bibfnamefont {D.}~\bibnamefont {Matthias}}, \bibinfo
  {author} {\bibfnamefont {J.}~\bibnamefont {Nunley}},\ and\ \bibinfo {author}
  {\bibfnamefont {A.}~\bibnamefont {Scheel}},\ }\href
  {https://doi.org/10.1103/PhysRevE.94.022219} {\bibfield  {journal} {\bibinfo
  {journal} {Physical Review E}\ }\textbf {\bibinfo {volume} {94}},\ \bibinfo
  {pages} {022219} (\bibinfo {year} {2016})},\ \bibinfo {note} {publisher:
  American Physical Society}\BibitemShut {NoStop}%
\bibitem [{{\relax DLMF}()}]{NIST}%
  \BibitemOpen
  {\relax DLMF},\ \href {https://dlmf.nist.gov/} {\bibinfo {title} {{\it NIST
  Digital Library of Mathematical Functions}}},\ \bibinfo {howpublished}
  {\url{https://dlmf.nist.gov/}, Release 1.1.10 of 2023-06-15},\ \bibinfo
  {note} {{F}.~W.~J. Olver, A.~B. {Olde Daalhuis}, D.~W. Lozier, B.~I.
  Schneider, R.~F. Boisvert, C.~W. Clark, B.~R. Miller, B.~V. Saunders, H.~S.
  Cohl, and M.~A. McClain, eds.}\BibitemShut {Stop}%
\bibitem [{\citenamefont {Ma}\ and\ \citenamefont
  {Knobloch}(2012)}]{ma_depinning_2012}%
  \BibitemOpen
  \bibfield  {author} {\bibinfo {author} {\bibfnamefont {Y.-P.}\ \bibnamefont
  {Ma}}\ and\ \bibinfo {author} {\bibfnamefont {E.}~\bibnamefont {Knobloch}},\
  }\href {https://doi.org/10.1063/1.4731268} {\bibfield  {journal} {\bibinfo
  {journal} {Chaos: An Interdisciplinary Journal of Nonlinear Science}\
  }\textbf {\bibinfo {volume} {22}},\ \bibinfo {pages} {033101} (\bibinfo
  {year} {2012})}\BibitemShut {NoStop}%
\bibitem [{\citenamefont {Asch}\ \emph {et~al.}(2023)\citenamefont {Asch},
  \citenamefont {Avery}, \citenamefont {Cortez},\ and\ \citenamefont
  {Scheel}}]{asch_slow_2023}%
  \BibitemOpen
  \bibfield  {author} {\bibinfo {author} {\bibfnamefont {A.}~\bibnamefont
  {Asch}}, \bibinfo {author} {\bibfnamefont {M.}~\bibnamefont {Avery}},
  \bibinfo {author} {\bibfnamefont {A.}~\bibnamefont {Cortez}},\ and\ \bibinfo
  {author} {\bibfnamefont {A.}~\bibnamefont {Scheel}},\ }\href
  {https://arxiv.org/abs/2309.14959v1} {\bibinfo {title} {Slow passage through
  the {Busse} balloon -- predicting steps on the {Eckhaus} staircase}}
  (\bibinfo {year} {2023}),\ \bibinfo {note} {preprint}\BibitemShut {NoStop}%
\bibitem [{\citenamefont {Knobloch}\ and\ \citenamefont
  {Merryfield}(1992)}]{knobloch_enhancement_1992}%
  \BibitemOpen
  \bibfield  {author} {\bibinfo {author} {\bibfnamefont {E.}~\bibnamefont
  {Knobloch}}\ and\ \bibinfo {author} {\bibfnamefont {W.~J.}\ \bibnamefont
  {Merryfield}},\ }\href {https://doi.org/10.1086/172052} {\bibfield  {journal}
  {\bibinfo  {journal} {The Astrophysical Journal}\ }\textbf {\bibinfo {volume}
  {401}},\ \bibinfo {pages} {196} (\bibinfo {year} {1992})}\BibitemShut
  {NoStop}%
\bibitem [{\citenamefont {Farrell}\ and\ \citenamefont
  {Ioannou}(1996)}]{farrell_generalized_1996}%
  \BibitemOpen
  \bibfield  {author} {\bibinfo {author} {\bibfnamefont {B.~F.}\ \bibnamefont
  {Farrell}}\ and\ \bibinfo {author} {\bibfnamefont {P.~J.}\ \bibnamefont
  {Ioannou}},\ }\href
  {https://doi.org/10.1175/1520-0469(1996)053<2041:GSTPIN>2.0.CO;2} {\bibfield
  {journal} {\bibinfo  {journal} {Journal of the Atmospheric Sciences}\
  }\textbf {\bibinfo {volume} {53}},\ \bibinfo {pages} {2041} (\bibinfo {year}
  {1996})}\BibitemShut {NoStop}%
\bibitem [{\citenamefont {Tribelsky}\ \emph {et~al.}(1992)\citenamefont
  {Tribelsky}, \citenamefont {Kai},\ and\ \citenamefont
  {Yamazaki}}]{tribelsky_phase-slip_1992}%
  \BibitemOpen
  \bibfield  {author} {\bibinfo {author} {\bibfnamefont {M.~I.}\ \bibnamefont
  {Tribelsky}}, \bibinfo {author} {\bibfnamefont {S.}~\bibnamefont {Kai}},\
  and\ \bibinfo {author} {\bibfnamefont {H.}~\bibnamefont {Yamazaki}},\ }\href
  {https://doi.org/10.1103/PhysRevA.45.4175} {\bibfield  {journal} {\bibinfo
  {journal} {Physical Review A}\ }\textbf {\bibinfo {volume} {45}},\ \bibinfo
  {pages} {4175} (\bibinfo {year} {1992})}\BibitemShut {NoStop}%
\bibitem [{\citenamefont {Uecker}(2014)}]{uecker_pde2path_2014}%
  \BibitemOpen
  \bibfield  {author} {\bibinfo {author} {\bibfnamefont {H.}~\bibnamefont
  {Uecker}},\ }\href {https://doi.org/10.4208/nmtma.2014.1231nm} {\bibfield
  {journal} {\bibinfo  {journal} {Numerical Mathematics: Theory, Methods and
  Applications}\ }\textbf {\bibinfo {volume} {7}},\ \bibinfo {pages} {58}
  (\bibinfo {year} {2014})}\BibitemShut {NoStop}%
\bibitem [{\citenamefont {Uecker}(2021{\natexlab{a}})}]{uecker2021numerical}%
  \BibitemOpen
  \bibfield  {author} {\bibinfo {author} {\bibfnamefont {H.}~\bibnamefont
  {Uecker}},\ }\href {https://doi.org/https://doi.org/10.1137/1.9781611976618}
  {\emph {\bibinfo {title} {{Numerical Continuation and Bifurcation in
  Nonlinear PDEs}}}}\ (\bibinfo  {publisher} {SIAM},\ \bibinfo {year}
  {2021})\BibitemShut {NoStop}%
\bibitem [{\citenamefont {Weideman}\ and\ \citenamefont
  {Reddy}(2000)}]{weideman2000matlab}%
  \BibitemOpen
  \bibfield  {author} {\bibinfo {author} {\bibfnamefont {J.~A.}\ \bibnamefont
  {Weideman}}\ and\ \bibinfo {author} {\bibfnamefont {S.~C.}\ \bibnamefont
  {Reddy}},\ }\href {https://doi.org/https://doi.org/10.1145/365723.365727}
  {\bibfield  {journal} {\bibinfo  {journal} {ACM Transactions on Mathematical
  Software}\ }\textbf {\bibinfo {volume} {26}},\ \bibinfo {pages} {465}
  (\bibinfo {year} {2000})}\BibitemShut {NoStop}%
\bibitem [{\citenamefont {Uecker}(2021{\natexlab{b}})}]{uecker2021pde2path}%
  \BibitemOpen
  \bibfield  {author} {\bibinfo {author} {\bibfnamefont {H.}~\bibnamefont
  {Uecker}},\ }\href
  {http://www.staff.uni-oldenburg.de/hannes.uecker/pde2path/tuts/modtut.pdf}
  {\bibinfo {title} {{pde2path without finite elements}}} (\bibinfo {year}
  {2021}{\natexlab{b}})\BibitemShut {NoStop}%
\bibitem [{\citenamefont {Liu}\ \emph {et~al.}(2022)\citenamefont {Liu},
  \citenamefont {Julien},\ and\ \citenamefont {Knobloch}}]{liu2022staircase}%
  \BibitemOpen
  \bibfield  {author} {\bibinfo {author} {\bibfnamefont {C.}~\bibnamefont
  {Liu}}, \bibinfo {author} {\bibfnamefont {K.}~\bibnamefont {Julien}},\ and\
  \bibinfo {author} {\bibfnamefont {E.}~\bibnamefont {Knobloch}},\ }\href
  {https://doi.org/https://doi.org/10.1017/jfm.2022.865} {\bibfield  {journal}
  {\bibinfo  {journal} {Journal of Fluid Mechanics}\ }\textbf {\bibinfo
  {volume} {952}},\ \bibinfo {pages} {A4} (\bibinfo {year} {2022})}\BibitemShut
  {NoStop}%
\bibitem [{\citenamefont {Liu}\ and\ \citenamefont
  {Knobloch}(2022)}]{liu2022single}%
  \BibitemOpen
  \bibfield  {author} {\bibinfo {author} {\bibfnamefont {C.}~\bibnamefont
  {Liu}}\ and\ \bibinfo {author} {\bibfnamefont {E.}~\bibnamefont {Knobloch}},\
  }\href {https://doi.org/https://doi.org/10.3390/fluids7120373} {\bibfield
  {journal} {\bibinfo  {journal} {Fluids}\ }\textbf {\bibinfo {volume} {7}},\
  \bibinfo {pages} {373} (\bibinfo {year} {2022})}\BibitemShut {NoStop}%
\bibitem [{\citenamefont {Liu}\ \emph {et~al.}(2024)\citenamefont {Liu},
  \citenamefont {Sharma}, \citenamefont {Julien},\ and\ \citenamefont
  {Knobloch}}]{liu2023fixed}%
  \BibitemOpen
  \bibfield  {author} {\bibinfo {author} {\bibfnamefont {C.}~\bibnamefont
  {Liu}}, \bibinfo {author} {\bibfnamefont {M.}~\bibnamefont {Sharma}},
  \bibinfo {author} {\bibfnamefont {K.}~\bibnamefont {Julien}},\ and\ \bibinfo
  {author} {\bibfnamefont {E.}~\bibnamefont {Knobloch}},\ }\href
  {https://doi.org/https://doi.org/10.1017/jfm.2023.1057} {\bibfield  {journal}
  {\bibinfo  {journal} {Journal of Fluid Mechanics}\ }\textbf {\bibinfo
  {volume} {979}},\ \bibinfo {pages} {A19} (\bibinfo {year}
  {2024})}\BibitemShut {NoStop}%
\bibitem [{\citenamefont {Burns}\ \emph {et~al.}(2020)\citenamefont {Burns},
  \citenamefont {Vasil}, \citenamefont {Oishi}, \citenamefont {Lecoanet},\ and\
  \citenamefont {Brown}}]{burns_dedalus_2020}%
  \BibitemOpen
  \bibfield  {author} {\bibinfo {author} {\bibfnamefont {K.~J.}\ \bibnamefont
  {Burns}}, \bibinfo {author} {\bibfnamefont {G.~M.}\ \bibnamefont {Vasil}},
  \bibinfo {author} {\bibfnamefont {J.~S.}\ \bibnamefont {Oishi}}, \bibinfo
  {author} {\bibfnamefont {D.}~\bibnamefont {Lecoanet}},\ and\ \bibinfo
  {author} {\bibfnamefont {B.~P.}\ \bibnamefont {Brown}},\ }\href
  {https://doi.org/10.1103/PhysRevResearch.2.023068} {\bibfield  {journal}
  {\bibinfo  {journal} {Physical Review Research}\ }\textbf {\bibinfo {volume}
  {2}},\ \bibinfo {pages} {023068} (\bibinfo {year} {2020})}\BibitemShut
  {NoStop}%
\bibitem [{\citenamefont {Ebert}\ and\ \citenamefont
  {Van~Saarloos}(2000)}]{ebert_front_2000}%
  \BibitemOpen
  \bibfield  {author} {\bibinfo {author} {\bibfnamefont {U.}~\bibnamefont
  {Ebert}}\ and\ \bibinfo {author} {\bibfnamefont {W.}~\bibnamefont
  {Van~Saarloos}},\ }\href {https://doi.org/10.1016/S0167-2789(00)00068-3}
  {\bibfield  {journal} {\bibinfo  {journal} {Physica D: Nonlinear Phenomena}\
  }\textbf {\bibinfo {volume} {146}},\ \bibinfo {pages} {1} (\bibinfo {year}
  {2000})}\BibitemShut {NoStop}%
\bibitem [{\citenamefont {Aronson}\ and\ \citenamefont
  {Weinberger}(1978)}]{aronson_multidimensional_1978}%
  \BibitemOpen
  \bibfield  {author} {\bibinfo {author} {\bibfnamefont {D.~G.}\ \bibnamefont
  {Aronson}}\ and\ \bibinfo {author} {\bibfnamefont {H.~F.}\ \bibnamefont
  {Weinberger}},\ }\href {https://doi.org/10.1016/0001-8708(78)90130-5}
  {\bibfield  {journal} {\bibinfo  {journal} {Advances in Mathematics}\
  }\textbf {\bibinfo {volume} {30}},\ \bibinfo {pages} {33} (\bibinfo {year}
  {1978})}\BibitemShut {NoStop}%
\bibitem [{\citenamefont {Ben-Jacob}\ \emph {et~al.}(1985)\citenamefont
  {Ben-Jacob}, \citenamefont {Brand}, \citenamefont {Dee}, \citenamefont
  {Kramer},\ and\ \citenamefont {Langer}}]{ben-jacob_pattern_1985}%
  \BibitemOpen
  \bibfield  {author} {\bibinfo {author} {\bibfnamefont {E.}~\bibnamefont
  {Ben-Jacob}}, \bibinfo {author} {\bibfnamefont {H.}~\bibnamefont {Brand}},
  \bibinfo {author} {\bibfnamefont {G.}~\bibnamefont {Dee}}, \bibinfo {author}
  {\bibfnamefont {L.}~\bibnamefont {Kramer}},\ and\ \bibinfo {author}
  {\bibfnamefont {J.~S.}\ \bibnamefont {Langer}},\ }\href
  {https://doi.org/10.1016/0167-2789(85)90094-6} {\bibfield  {journal}
  {\bibinfo  {journal} {Physica D: Nonlinear Phenomena}\ }\textbf {\bibinfo
  {volume} {14}},\ \bibinfo {pages} {348} (\bibinfo {year} {1985})}\BibitemShut
  {NoStop}%
\bibitem [{\citenamefont {Dee}(1985)}]{dee_dynamical_1985}%
  \BibitemOpen
  \bibfield  {author} {\bibinfo {author} {\bibfnamefont {G.}~\bibnamefont
  {Dee}},\ }\href {https://doi.org/10.1016/S0167-2789(85)80001-4} {\bibfield
  {journal} {\bibinfo  {journal} {Physica D: Nonlinear Phenomena}\ }\textbf
  {\bibinfo {volume} {15}},\ \bibinfo {pages} {295} (\bibinfo {year}
  {1985})}\BibitemShut {NoStop}%
\bibitem [{\citenamefont {Gelens}\ and\ \citenamefont
  {Knobloch}(2010)}]{gelens_coarsening_2010}%
  \BibitemOpen
  \bibfield  {author} {\bibinfo {author} {\bibfnamefont {L.}~\bibnamefont
  {Gelens}}\ and\ \bibinfo {author} {\bibfnamefont {E.}~\bibnamefont
  {Knobloch}},\ }\href {https://doi.org/10.1140/epjd/e2010-00132-6} {\bibfield
  {journal} {\bibinfo  {journal} {The European Physical Journal D}\ }\textbf
  {\bibinfo {volume} {59}},\ \bibinfo {pages} {23} (\bibinfo {year}
  {2010})}\BibitemShut {NoStop}%
\bibitem [{\citenamefont {Hoyle}(2006)}]{hoyle_pattern_2006}%
  \BibitemOpen
  \bibfield  {author} {\bibinfo {author} {\bibfnamefont {R.}~\bibnamefont
  {Hoyle}},\ }\href {https://doi.org/10.1017/CBO9780511616051} {\emph {\bibinfo
  {title} {Pattern {Formation}: {An} {Introduction} to {Methods}}}}\ (\bibinfo
  {publisher} {Cambridge University Press},\ \bibinfo {address} {Cambridge},\
  \bibinfo {year} {2006})\BibitemShut {NoStop}%
\bibitem [{\citenamefont {Collet}\ and\ \citenamefont
  {Eckmann}(1992)}]{collet_solutions_1992}%
  \BibitemOpen
  \bibfield  {author} {\bibinfo {author} {\bibfnamefont {P.}~\bibnamefont
  {Collet}}\ and\ \bibinfo {author} {\bibfnamefont {J.~P.}\ \bibnamefont
  {Eckmann}},\ }\href {https://doi.org/10.1007/BF02099141} {\bibfield
  {journal} {\bibinfo  {journal} {Communications in Mathematical Physics}\
  }\textbf {\bibinfo {volume} {145}},\ \bibinfo {pages} {345} (\bibinfo {year}
  {1992})}\BibitemShut {NoStop}%
\bibitem [{\citenamefont {Moehlis}\ and\ \citenamefont
  {Knobloch}(1996)}]{moehlis_eckhaus-benjamin-feir_1996}%
  \BibitemOpen
  \bibfield  {author} {\bibinfo {author} {\bibfnamefont {J.}~\bibnamefont
  {Moehlis}}\ and\ \bibinfo {author} {\bibfnamefont {E.}~\bibnamefont
  {Knobloch}},\ }\href {https://doi.org/10.1103/PhysRevE.54.5161} {\bibfield
  {journal} {\bibinfo  {journal} {Physical Review E}\ }\textbf {\bibinfo
  {volume} {54}},\ \bibinfo {pages} {5161} (\bibinfo {year}
  {1996})}\BibitemShut {NoStop}%
\bibitem [{\citenamefont {Ponedel}\ \emph {et~al.}(2017)\citenamefont
  {Ponedel}, \citenamefont {Kao},\ and\ \citenamefont
  {Knobloch}}]{ponedel_front_2017}%
  \BibitemOpen
  \bibfield  {author} {\bibinfo {author} {\bibfnamefont {B.~C.}\ \bibnamefont
  {Ponedel}}, \bibinfo {author} {\bibfnamefont {H.-C.}\ \bibnamefont {Kao}},\
  and\ \bibinfo {author} {\bibfnamefont {E.}~\bibnamefont {Knobloch}},\ }\href
  {https://doi.org/10.1103/PhysRevE.96.032208} {\bibfield  {journal} {\bibinfo
  {journal} {Physical Review E}\ }\textbf {\bibinfo {volume} {96}},\ \bibinfo
  {pages} {032208} (\bibinfo {year} {2017})}\BibitemShut {NoStop}%
\bibitem [{\citenamefont {Aranson}\ and\ \citenamefont
  {Kramer}(2002)}]{aranson_world_2002}%
  \BibitemOpen
  \bibfield  {author} {\bibinfo {author} {\bibfnamefont {I.~S.}\ \bibnamefont
  {Aranson}}\ and\ \bibinfo {author} {\bibfnamefont {L.}~\bibnamefont
  {Kramer}},\ }\href {https://doi.org/10.1103/RevModPhys.74.99} {\bibfield
  {journal} {\bibinfo  {journal} {Reviews of Modern Physics}\ }\textbf
  {\bibinfo {volume} {74}},\ \bibinfo {pages} {99} (\bibinfo {year}
  {2002})}\BibitemShut {NoStop}%
\end{thebibliography}
\end{document}